\documentstyle[12pt,epsfig]{article}
\def\thebibliography#1{\centerline{\bf REFERENCES}\list
  {[\arabic{enumi}]}{\settowidth\labelwidth{[#1]}\leftmargin\labelwidth
    \advance\leftmargin\labelsep
    \usecounter{enumi}}
    \def\newblock{\hskip .11em plus .33em minus .07em}
    \sloppy\clubpenalty4000\widowpenalty4000}

\newcommand{\csi}{{\rm cos}(\theta^\pm_p)}
\newcommand{\csii}{{\rm cos}^2(\theta^\pm_p)}
\newcommand{\gvp}{\tilde{g}_{\rm v}\left(p\right)}
\newcommand{\fvp}{\tilde{f}_{\rm v}\left(p\right)}
\newcommand{\gip}{\tilde{g}^{\left(1\right)}\left(p\right)}
\newcommand{\fip}{\tilde{f}^{\left(1\right)}\left(p\right)}
\newcommand{\giip}{\tilde{g}^{\left(2\right)}\left(p\right)}
\newcommand{\fiip}{\tilde{f}^{\left(2\right)}\left(p\right)}
\newcommand{\fup}{\tilde{f}_\mu^{\left(i\right)}\left(p\right)}
\newcommand{\fupq}{\tilde{f}^{\left(i\right)}\left(p\right)}
\newcommand{\gup}{\tilde{g}_\mu^{\left(i\right)}\left(p\right)}
\newcommand{\gupq}{\tilde{g}^{\left(i\right)}\left(p\right)}
\newcommand{\guip}{\tilde{g}_{\mu}^{\left(1\right)}\left(p\right)}
\newcommand{\fuip}{\tilde{f}_{\mu}^{\left(1\right)}\left(p\right)}
\newcommand{\guiip}{\tilde{g}_{\mu}^{\left(2\right)}\left(p\right)}
\newcommand{\fuiip}{\tilde{f}_{\mu}^{\left(2\right)}\left(p\right)}

\newcommand{\gvr}{g_{\rm v}\left(r\right)}
\newcommand{\fvr}{f_{\rm v}\left(r\right)}
\newcommand{\guiir}{g_{\mu}^{\left(2\right)}\left(r\right)}
\newcommand{\fuiir}{f_{\mu}^{\left(2\right)}\left(r\right)}
\newcommand{\be}{\begin{eqnarray}}
\newcommand{\ba}{\begin{array}}
\newcommand{\ea}{\end{array}}
\newcommand{\ee}{\end{eqnarray}}

\newcommand{\dslash}{\partial \hskip -0.5em /}

\newcommand{\Tr}{{\rm Tr}}

\newcommand{\La}{{\cal L}}
\newcommand{\A}{{\cal A}}

\newcommand{\bbox}[1]{\mbox{\boldmath$#1$\unboldmath}}

\newcommand{\bfxi}{\mbox{\scriptsize\boldmath $\xi$}}
\newcommand{\zr}[1]{\mbox{\hspace*{#1em}}}
\newcommand{\ID}{\mbox{{\sf 1}\zr{-0.16}\rule{0.04em}{1.55ex}\zr{0.1}}}

\begin{document}
\rightline{January (revised May) 1998}
\rightline{OKHEP-97-07}
\rightline{UNITU-THEP-1/1998}
\rightline{hep-ph/9801379}
\vskip 1.0truecm
\centerline{\Large\bf Chiral Odd Structure Functions from a Chiral Soliton}
\baselineskip=16 true pt
\vskip 1.0cm
\centerline{L.\ Gamberg$^{a)}$, H.\ Reinhardt$^{b)}$
and H.\ Weigel$^{b)}$}
\vskip 0.5cm
\centerline{$^{a)}$Department of Physics and Astronomy}
\centerline{University of Oklahoma}
\centerline{440 W. Brooks Ave}
\centerline{Norman, OK 73019--0225, USA}
\vskip 0.5cm
\centerline{$^{b)}$Institute for Theoretical Physics}
\centerline{T\"ubingen University}
\centerline{Auf der Morgenstelle 14}
\centerline{D-72076 T\"ubingen, Germany}
\vskip 1.0cm
\baselineskip=16pt
\centerline{\bf ABSTRACT}
\bigskip
We calculate the chiral odd quark distributions and the 
corresponding structure functions $h_T(x,Q^2)$ and
$h_L(x,Q^2)$ within the Nambu--Jona--Lasinio chiral soliton
model for the nucleon.  The $Q^2$ evolution of the twist--2
contributions is performed according to the standard
GLAP formalism while the twist--three piece, $\overline{h}_L(x)$,
is evolved according to the large $N_C$ scheme. We carry out
a comparison between the chiral odd structure functions of the
proton and the neutron. At the low model scale ($Q_0^2$) we find 
that the leading twist effective quark distributions, 
$f_1^{(q)}(x,Q_0^2)$, $g_1^{(q)}(x,Q_0^2)$ and $h_T^{(q)}(x,Q_0^2)$ 
satisfy Soffer's inequality for both quark flavors 
$q=u,d$.

\vskip 2.0cm
\leftline{\it PACS: 11.30.Cp, 12.39.Ki.}
\vskip 1.0cm
\leftline{To be published in: Phys. Rev. {\bf D}.}
\newpage

\normalsize\baselineskip=20pt
\bigskip
\section{Introduction}
\bigskip
There have been a number of recent investigations into the 
{\em chiral odd} structure functions of the nucleon. As in the case 
of the polarized structure functions there are two quantities of 
interest at leading twist: The transverse spin chiral odd structure
function $h_T(x,Q^2)$ and the longitudinal spin chiral odd structure 
function $h_L(x,Q^2)$. Within the context of the operator product 
expansion (OPE) the analysis in terms of twist reveals that the 
transverse chiral odd structure function $h_T(x,Q^2)$ is purely twist--2, 
while  the longitudinal structure function $h_L(x,Q^2)$ contains 
both twist--2 and twist--3 contributions.
Accordingly, the  decomposition of $h_L(x,Q^2)$ into twist--2 
and twist--3 ($\overline{h}_L(x,Q^2)$) pieces is given by
\be
h_L(x,Q^2)=2x\, \int_{x}^1\ dy\frac{h_T(y,Q^2)}{y^2}+ 
\overline{h}_L(x,Q^2)\ .
\label{hL}
\ee
As a reminder we note that the kinematics are defined such that $q$ denotes 
the momentum transferred to a nucleon of momentum $p$. In the Bjorken
limit, {\it i.e.} $Q^2=-q^2\to\infty$ with $x=Q^2/2p\cdot\! q$ fixed,
the leading twist contributions to the nucleon structure
functions dominate the $1/Q^2$ expansion. The additional and 
important logarithmic dependence on $Q^2$, which is associated 
with soft gluon emission, is included via the evolution program 
of perturbative quantum--chromo--dynamics (QCD).

While the chiral odd structure functions are not directly accessible 
in deep inelastic lepton nucleon scattering (DIS) there is the well known 
proposal at {\em RHIC} to extract the quark transversality distributions
$h_T^{(a)}(x,Q^2)$ ($a$ being the flavor index) from Drell--Yan 
dilepton--production resulting from transversely polarized proton 
beams \cite{Ra79}. Unfortunately dilepton production processes 
are difficult to extract from proton--proton collisions as the purely
hadronic processes dominate. Furthermore this experiment will provide
only the product of the chiral odd distributions for quarks and 
antiquarks. As the latter are presumably small these flavor distributions 
are not easily measurable in the Drell--Yan process. In the light of these 
disadvantages it has recently been pointed out that the transversality 
distributions may also be measured in the fragmentation region of 
DIS \cite{Ja98}. The key observation is that these distribution 
functions can be extracted from an asymmetry in the two meson production 
in the special case that this two meson state (like $\pi^+\pi^-$) is a 
superposition of different $C$--parity states, as {\it e.g.} $\sigma$ 
and $\rho$. Then the phases in the final state interactions do not 
vanish on the average and the differential cross section is proportional 
to the product of chiral odd distributions and the interference 
fragmentation functions. The latter describe the emission and 
subsequent absorption of a two pion intermediate state from quarks 
of different helicity. In case these fragmentation functions are not 
anomalously small the chiral odd distribution functions can then be 
obtained from DIS processes\footnote{The relevant fragmentation and 
distribution functions depend on different kinematical variables: the 
two meson state momentum fraction and the Bjorken variable, respectively.} 
like $eN\to e^\prime \pi^+\pi^-X$ with the nucleon $N$ being transversely 
polarized. Assuming isospin covariance for the fragmentation functions 
these DIS processes will provide access to the charge squared weighted 
chiral odd distribution functions \cite{Ja98}. Such processes 
should be measurable in the transversely polarized target experiments at 
{\it HERMES}. Knowledge of the chiral odd structure 
functions will serve to complete our picture of the spin structure
of the nucleon as they correspond to the distribution of the quark 
transverse spin in a nucleon which is transversely polarized 
\cite{Ja96}. With these data being expected in the near future 
it is, of course, interesting to understand the structure of 
the nucleon from the theoretical point of view. As we are still 
lacking a bound state wave function for nucleon in terms of quarks 
and gluons, {\it i.e.} computed from first principles in QCD,
it is both mandatory and fruitful to investigate these chiral odd  
flavor distributions and their charge weighted average
nucleon structure functions within hadronic models of the
nucleon \cite{Ja92,St93,Ba97,Sc97,Sc97a,Ka97,Po96}. 

In the context of the spin structure of the nucleon chiral soliton
models are particularly interesting as they provide an  explanation
for the small magnitude of the quark spin contribution to the proton 
spin, {\it i.e.} the vanishingly small matrix element of the singlet 
axial current \cite{We96}. In these models the nucleon is described as a 
non--perturbative field configuration in some non--linear effective 
meson theory \cite{Sk61,Ad83,Al96}. Unfortunately in many of these soliton 
models the evaluation of structure functions is infeasible due to the 
highly non--linear structure of the current operators and the inclusion 
of higher derivative operators which complicates the current commutation 
relations. However, it has recently been recognized that the soliton 
solution \cite{Al96} which emerges after bosonization \cite{Eb86} of 
the Nambu--Jona--Lasinio (NJL) \cite{Na61} chiral quark model can be 
employed to compute nucleon structure functions \cite{We96a,We97}. In 
order to project this soliton configuration onto nucleon states with good 
spin and flavor a cranking procedure must be employed \cite{Ad83,Re89} 
which implements significant $1/N_C$ contributions ($N_C$ is the number 
of color degrees of freedom.). When extracting the structure functions 
from the NJL chiral soliton model the full calculation which also
includes effects of the vacuum polarized by the background soliton is 
quite laborious. In addition we are still lacking a regularization 
prescription of the vacuum contribution to the structure functions
which is derived from the action functional and which yields algebraic 
expressions for their moments which are {\em consistent} with those for 
the static nucleon properties. Fortunately it is known that the 
dominant contribution to 
static nucleon properties stems from the single quark level which has the 
lowest energy eigenvalue (in magnitude) and is strongly bound by the 
soliton \cite{Al96}. This is particularly the case for spin related 
quantities. Hence it is a reasonable approximation to consider
only the contribution of this level to the structure functions. In 
the proceeding section the NJL chiral soliton model together with the 
above mentioned approximation, which we will call {\it valence quark 
approximation}\footnote{This notation refers to the valence quark in the 
NJL chiral soliton model and should not be confused with the valence quark
in the parton model.} will be described in more detail. 

The NJL model for the quark flavor dynamics incorporates spontaneous
breaking of chiral symmetry in a dynamic fashion. Hence the quark fields 
which built up the soliton self--consistently \cite{Re88} are {\em 
constituent quarks} with a constituent quark mass of several hundred 
{\rm MeV}. Keeping this in mind we calculate both the {\em effective}
constituent quark distributions and in turn the corresponding leading twist
contributions to nucleon structure functions ({\it cf.} eq (\ref{chgw}))
at a low scale $Q_0^2$. In the language of Feynman diagrams 
the DIS processes are described by a constituent quark of the nucleon 
absorbing a quanta of the external source. In the Bjorken limit the quark 
then propagates highly off--shell before emitting a quanta of the external 
source. The intermediate quark may propagate forward and backward.
Hence the complete structure functions acquire contributions from 
both distributions where the intermediate constituent quark moves 
forward and backward. We will focus on nucleon structure functions which
are defined as the sum over the charge--weighted flavor distributions 
\cite{Ja92}
\be
h_{T/L}^{(\pm)}(x,Q_0^2)=\frac{1}{2} \sum_a  
e_{a}^2 h^{(a,\pm)}_{T/L}(x,Q_0^2) \ ,
\label{chgw}
\ee
in analogy to those of the chiral even spin polarized and unpolarized 
nucleon structure functions \cite{Ja98,Ja96}. Here $a$ represents a 
quark label, while $(\pm)$ refers to the forward $(+)$ and backward
$(-)$ propagating intermediate constituent quarks. Furthermore $e_{a}$ 
denotes the charge fraction of the considered quark flavor $a$. The 
complete chiral odd structure functions are finally obtained as the sum
\be
h_{T/L}(x,Q_0^2)=h_{T/L}^{(+)}(x,Q_0^2)+h_{T/L}^{(-)}(x,Q_0^2)\ .
\label{defintro}
\ee
The calculation of the flavor distributions $h^{(a)}_{T/L}$ in the 
valence approximation to the NJL chiral soliton model \cite{We96a,We97} 
is summarized in section 3.

Further it is important to note that when considering model structure 
functions the OPE implies that the initial conditions,
$\mu^2=Q_0^2$, for the evolution is, 
{\it a priori}, a free parameter in any baryon model \cite{Sc91}. 
For the model under consideration it has previously been determined to 
$Q_0^2\approx0.4{\rm GeV}^2$ by studying the evolution dependence of 
the model prediction for the unpolarized structure functions \cite{We96a}. 
In a subsequent step to compute the chiral odd structure functions we 
employ a leading order evolution program \cite{Ba97,Ka97} to obtain the 
chiral odd structure functions at a larger scale, {\it e.g.} 
$Q^2\approx 4{\rm GeV}^2$ relevant to the experimental conditions. This 
evolution program incorporates the leading logarithmic corrections to 
the leading twist pieces. The evolution procedure as applied to our 
model structure functions will be explained in section 4.

The numerical results for the chiral odd structure functions are
presented in section 5 while concluding remarks are contained in 
section 6. Technical details on the model calculations and the QCD 
evolution procedure are relegated to appendices. Let us also mention
that there has been a previous calculation of $h_T(x,Q_0^2)$ \cite{Po96} 
which, however, ignored both the projection onto good nucleon states 
and the QCD evolution. Furthermore in that calculation an (arbitrary) 
meson profile was employed rather than a self--consistent soliton 
solution to the static equations of motion. 

\bigskip
\section{The NJL--Model Chiral Soliton}
\bigskip
 
Before continuing with the discussion of the chiral odd structure
functions, we will review the issue of the 
chiral soliton in the NJL model.

The Lagrangian of the NJL model in terms of quark degrees of freedom 
reads \cite{Na61,Eb86}
\be
\La = \bar q (i\dslash -  m^0 ) q +
      2G_{\rm NJL} \sum _{i=0}^{3}
\left( (\bar q \frac {\tau^i}{2} q )^2
      +(\bar q \frac {\tau^i}{2} i\gamma _5 q )^2 \right) .
\label{NJL}
\ee
Here $q$, $\hat m^0$ and $G_{\rm NJL}$ denote the quark field, the 
current quark mass and a dimensionful coupling constant, respectively.
This model is motivated as follows:
Integrating out the gluon fields from QCD yields a current--current 
interaction mediated by one gluon exchange to leading order
in powers of the quark current.  Replacing the gluon mediating 
propagator with a local contact interaction and 
performing the appropriate Fierz--transformations yields the 
Lagrangian (\ref{NJL}) in leading order of $1/N_C$ \cite{Ca87,Re90}, 
where $N_C$ refers to the number of color degrees of freedom. Although
only a subset of possible non--perturbative gluonic modes are 
contained in the contact interaction term in eq (\ref{NJL}) 
it is important to stress that gluonic effects are contained in the 
model (\ref{NJL}). Furthermore the NJL model embodies the approximate 
chiral symmetry of QCD and has to be understood as an effective 
(non--renormalizable) theory of the low--energy quark flavor dynamics.

Application of functional bosonization techniques \cite{Eb86} to the 
Lagrangian (\ref{NJL}) yields the mesonic action
\be
\A&=&\Tr_\Lambda\log(D)+\frac{1}{4G_{\rm NJL}}
\int d^4x\ {\rm tr}
\left(m^0\left(M+M^{\dag}\right)-MM^{\dag}\right)\ , 
\label{bosact} \\
D&=&i\dslash-\left(M+M^{\dag}\right)
-\gamma_5\left(M-M^{\dag}\right)\ ,
\label{dirac}
\ee
where $M=S+iP$ comprises composite scalar ($S$) and pseudoscalar ($P$) 
meson fields which appear as quark--antiquark bound states.  
For regularization, which is indicated by the cut--off $\Lambda$, we 
will adopt the proper--time scheme \cite{Sch51}. The free parameters 
of the model are the current quark mass $m^0$, the coupling constant 
$G_{\rm NJL}$ and the cut--off $\Lambda$. The equation of motion for 
the scalar field $S$ may be considered as the gap--equation for the 
order parameter $\langle {\bar q} q\rangle$ of chiral symmetry breaking. 
This equation relates the vacuum expectation value 
$\langle M\rangle=m{\ID}$ to the model parameters $m^0$, $G_{\rm NJL}$ 
and $\Lambda$. For apparent reasons $m$ is called the {\em constituent} 
quark mass. The occurrence of this vacuum expectation value reflects the 
spontaneous breaking of chiral symmetry and causes the pseudoscalar fields 
to emerge as (would--be) Goldstone bosons. Expanding $\A$ to quadratic 
order in $P$ (around $\langle M\rangle$) these parameters are related to 
physical quantities; that is, the pion mass, $m_\pi=135{\rm MeV}$ and the 
pion decay constant, $f_\pi=93{\rm MeV}$. This leaves one undetermined 
parameter which we choose to be the constituent quark mass \cite{Eb86}.

The NJL model chiral soliton \cite{Al96,Re88} is given 
by a non--perturbative meson configuration which is assumed of the 
hedgehog type
\be
M_{\rm H}(\mbox{\boldmath $x$})=m\ {\rm exp}
\left(i\mbox{\boldmath $\tau$}\cdot{\hat{\mbox{\boldmath $x$}}}
\Theta(r)\right)\ .
\label{hedgehog}
\ee
In order to compute the functional trace in eq (\ref{bosact}) for this 
static configuration we express the 
Dirac operator (\ref{dirac}) in terms of a Hamiltonian
operator $h$, {\it i.e.}  $D=i\beta(\partial_t-h)$ with
\be
h=\mbox{\boldmath $\alpha$}\cdot\mbox{\boldmath $p$}+m\ \beta\
{\rm exp}\left(i\gamma_5\mbox{\boldmath $\tau$}
\cdot{\hat{\mbox{\boldmath $x$}}}\Theta(r)\right)\ .
\label{hamil}
\ee
We denote the eigenvalues and eigenfunctions of $h$ by 
$\epsilon_\mu$ and $\Psi_\mu$, respectively. Explicit expressions for 
these wave--functions are displayed in appendix B of ref \cite{Al96}. 
In the proper--time regularization scheme the energy functional of 
the NJL model is found to be \cite{Re89,Al96}, 
\be
E[\Theta]=
\frac{N_C}{2}\epsilon_{\rm v}
\left(1+{\rm sgn}(\epsilon_{\rm v})\right)
&+&\frac{N_C}{2}\int^\infty_{1/\Lambda^2}
\frac{ds}{\sqrt{4\pi s^3}}\sum_\nu{\rm exp}
\left(-s\epsilon_\nu^2\right)
\nonumber \\* && \hspace{1.5cm}
+\ m_\pi^2 f_\pi^2\int d^3r  \left(1-{\rm cos}\Theta(r)\right) .
\label{efunct}
\ee
The subscript ``${\rm v}$" denotes the valence quark level. This state 
is the distinct level bound in the soliton background, {\it i.e.}
$-m<\epsilon_{\rm v}<m$. The chiral angle, $\Theta(r)$, is
obtained by self--consistently extremizing $E[\Theta]$ \cite{Re88}.

States possessing nucleon quantum numbers of spin and isospin are 
generated by elevating the rotational zero modes to time dependent 
large amplitude rotational fluctuations about the hedgehog field
\cite{Ad83}
\be
M(\mbox{\boldmath $x$},t)=
A(t)M_{\rm H}(\mbox{\boldmath $x$})A^{\dag}(t)\ ,
\label{collrot}
\ee
which introduces the collective coordinates $A(t)\in SU(2)$. 
Substituting the {\it ansatz} (\ref{collrot})
into the action functional (\ref{bosact}) and expanding
\cite{Re89} in the angular velocities 
\be 
2A^{\dag}(t)\dot A(t)=
i\mbox{\boldmath $\tau$}\cdot\mbox{\boldmath $\Omega$}
\label{angvel}
\ee
to quadratic order yields the Lagrange function for the collective 
coordinates.  Upon canonical quantization the angular velocity 
$\mbox{\boldmath $\Omega$}$ is substituted by the nucleon spin 
operator $\mbox{\boldmath $J$}=\alpha^2\mbox{\boldmath $\Omega$}$,
with $\alpha^2$ being the moment of inertia \cite{Re89,Al96}.
The eigenfunctions of the resulting Hamiltonian are the 
Wigner $D$--functions
\be
\langle A|N\rangle=\frac{1}{2\pi}
D^{1/2}_{I_3,-J_3}(A)\ ,
\label{nwfct}
\ee
with $I_3$ and $J_3$ being respectively the isospin and spin projection 
quantum numbers of the nucleon. The nucleon matrix elements of the 
collective rotations are obtained via
$\langle N | {\rm tr}(\tau_i A \tau_j A^\dagger) |N\rangle=
-(8/3)\langle N | I_i J_j |N\rangle$ \cite{Ad83}.
This approach to 
generate nucleon states from the hedgehog corresponds to the
cranking technique in nuclear physics \cite{In54}.

Expectation values of bilocal quark--bilinears appearing in the 
evaluation of nucleon structure functions are expressed as (regularized) 
sums over bilocal and bilinear combinations of all eigenfunctions 
$\Psi_\mu$ including the Dirac sea states. In practice this is quite 
a painful task, in particular when cranking corrections (\ref{collrot}) 
are included. Also the problem of regularization is not consistently 
solved. Fortunately it 
turns out that the dominant contributions ($\ge80\%$) to static nucleon 
properties (which are moments of the structure functions) stems from the 
distinct valence level $\Psi_{\rm v}$ \cite{Al96}. It is therefore 
reasonable to approximate the relevant bilinears by their valence quark 
contribution. In order to obtain quark distributions of the {\em nucleon}
and the corresponding nucleon structure functions ({\it c.f.} \ref{chgw}),
rather than {\em soliton} structure functions the cranking contribution
to the wave--function, which is induced by the collective rotation $A(t)$, 
must be included. That is, the valence quark wave--function employed to 
approximate the bilinears in the structure functions reads
\be
\Psi_{\rm v}(\mbox{\boldmath $x$},t)=
{\rm e}^{-i\epsilon_{\rm v}t}A(t)
\left\{\Psi_{\rm v}(\mbox{\boldmath $x$})
+\frac{1}{2}\sum_{\mu\ne{\rm v}}
\Psi_\mu(\mbox{\boldmath $x$})
\frac{\langle \mu |\mbox{\boldmath $\tau$}\cdot
\mbox{\boldmath $\Omega$}|{\rm v}\rangle}
{\epsilon_{\rm v}-\epsilon_\mu}\right\}=:
{\rm e}^{-i\epsilon_{\rm v}t}A(t)
\psi_{\rm v}(\mbox{\boldmath $x$}).
\label{valrot}
\ee
Here $\psi_{\rm v}(\mbox{\boldmath $x$})$ refers to the spatial part 
of the body--fixed valence quark wave--function with the rotational 
corrections included and 
$\Psi_\mu=\langle\mu|\mbox{\boldmath $x$}\rangle$ are eigenfunctions 
of the Dirac Hamiltonian (\ref{hamil}). This replacement of the bilocal 
and bilinear quark fields when computing nucleon structure functions defines 
the valence quark approximation.

\bigskip
\section{Chiral Odd Structure Functions, 
${\lowercase{h}}_T({\lowercase{x}})$ 
and ${\lowercase{h}}_L({\lowercase{x}})$
in the NJL model}
\bigskip

Here we present the major topic of this paper, namely the calculation 
of the twist--2 and twist--3 chiral odd structure functions in the 
NJL chiral soliton model. Like their deep inelastic chiral even 
(un)polarized counterparts, the chiral odd structure functions are
computed as Fourier transformations of nucleon matrix elements of  
bilocal quark operators on the light--cone \cite{Ja92}. The key 
features of the relevant light--cone kinematics are given in Appendix A.  

We begin by listing the forward propagating intermediate quark $(+)$ 
contribution to the chiral odd nucleon structure functions.
Before, however, straightforwardly transcribing the expressions from 
appendix A we must recall that the soliton represents a localized 
field configuration. Therefore a collective coordinate 
$\mbox{\boldmath $x$}_0$ is introduced which parameterizes the position 
of the soliton. This collective coordinate is employed to generate 
states with good linear momentum \cite{Ge75}. When computing matrix 
elements between states of identical momenta one is essentially left 
with an integration over $\mbox{\boldmath $x$}_0$. In the nucleon 
rest--frame (RF) the contribution of the forward moving intermediate 
quark to the chiral odd structure functions may therefore be 
expressed as\footnote{The following expressions constitute 
a generalization of Jaffe's original definition \cite{Ja75} for nucleon 
structure functions.}
\be
h^{(+)}_T(x)&=&N_C\frac{2M\sqrt{2}}{8\pi}\ \int d\xi^-
{\rm exp}(-i\xi^-\frac{Mx}{\sqrt{2}})
\nonumber \\ && \hspace{1.5cm}
\times\int d^3\bbox{x}_0\ 
\langle \bbox{S}_{\perp}|\Psi_+^\dagger(\xi - \bbox{x}_0) 
\gamma_\perp\gamma_5{\cal Q}^2\Psi_+(-\bbox{x}_0)
|\bbox{S}_\perp\Big\rangle_{\xi^+=\bfxi_\perp=0}\ .
\label{cht}
\ee
For convenience we have omitted the subscript ``v'' for the valence 
quark wave function. Note that $\xi$ refers to a four vector which 
in light--cone coordinates reads $(\xi^+,\xi^-,\bbox{\xi}_\perp)$.
This coordinate 
enters the light--cone variables via  $\xi^{\pm}=(t\pm z)/\sqrt{2}$. 
Also, the notation $\bbox{S}_{\perp}$ is synonymous for 
the spin being perpendicular to the coordinate $z$. 
On the other hand for the longitudinal counterpart
\be
h^{(+)}_L(x)&=&N_C\frac{2M\sqrt{2}}{16\pi}\ \int  d\xi^-
{\rm exp}(-i\xi^-\frac{Mx}{\sqrt{2}})
\nonumber \\ && \hspace{1.5cm}
\times\int d^3\bbox{x}_0\ 
\langle \bbox{S}_z|\Psi_+^\dagger(\xi - \bbox{x}_0) 
\gamma_0\gamma_5{\cal Q}^2\Psi_-(-\bbox{x}_0)
\nonumber \\ &&\hspace{4cm}
-\Psi_-^\dagger(\xi - \bbox{x}_0) 
\gamma_0\gamma_5{\cal Q}^2\Psi_+(-\bbox{x}_0)| 
\bbox{S}_z\rangle_{\xi^+=\bfxi_\perp=0}
\label{chl}
\ee
the spin is aligned with the $z$--axis.
The ``good'' and ``bad'' light--cone components of the quark wave 
functions are the projections $\Psi_{\pm}=P_{\pm}\Psi$, with
$P_{\pm}=\frac{1}{2}\gamma^{\mp}\gamma^{\pm}$ being the corresponding 
projections operators. Above, ${\cal Q}={\rm diag}(2/3,-1/3)$ refers 
to the matrix containing the quark charge fractions and the zero momentum 
nucleon states are given by, 
$|\bbox{p}=0,\bbox{S}\rangle
=\left[(2\pi)^3 2M\right]^{\frac{1}{2}}
|\bbox{S}\rangle$.
Introducing Fourier transforms for the spatial part of 
the valence quark wave functions ({\it cf.} eq (\ref{valrot})),
\be
\psi\left(\bbox{\xi}_\bot,
\xi_3=- \frac{\xi^-}{\scriptstyle\sqrt{2}}\right)=
\int \frac{d^2p_\bot dp_3}{2\pi^2}\
{\rm exp}\left[i\left(\frac{p_3\xi^-}{\scriptstyle\sqrt{2}}-
\bbox{p}_\bot\cdot
\bbox{\xi}_\bot\right)\right]
\tilde{\psi}\left(\bbox{p}_\bot,p_3\right) \ 
\label{ftrans}
\ee
yields,
\be
h^{(+)}_T(x)&=&N_C\frac{M}{\sqrt{2}\pi^2}\int d\xi^-\ 
p^2dp\ d({\rm cos}\, \theta)\ d\phi\ 
{\rm exp}\left(\frac{-i\xi^-
\left(Mx-{\rm\epsilon_{\rm v}}+p{\rm cos}\, \theta\right)}{\sqrt 2}\right)
\nonumber \\ &&\hspace{6cm}
\times\langle \bbox{S}_\perp |\tilde{\psi}_+^\dagger(\bbox{p})
\gamma_\perp\gamma_5{\cal Q}^2
\tilde{\psi}_+(\bbox{p})|\bbox{S}_\perp\rangle\ ,
\label{htpx}
\ee
and
\be
h^{(+)}_L(x)&=&N_C\frac{M}{\sqrt{2}\pi^2}\int d\xi^-\ 
p^2dp\ d({\rm cos}\, \theta)\ d\phi\ {\rm exp}\left(\frac{-i\xi^-
\left(Mx-{\rm\epsilon_{\rm v}}+p{\rm cos}\, \theta\right)}{\sqrt 2}\right)
\nonumber \\ &&\hspace{1cm}
\times\langle \bbox{S}_z |\tilde{\psi}_+^\dagger(\bbox{p})
\gamma_0\gamma_5{\cal Q}^2\tilde{\psi}_-(\bbox{p})
-\tilde{\psi}_-^\dagger(\bbox{p})
\gamma_0\gamma_5{\cal Q}^2\tilde{\psi}_+(\bbox{p})
|\bbox{S}_z\rangle\ .
\label{hlpx}
\ee
In anticipation of the decomposition (\ref{valrot}) the square of the 
charge operator is redefined to
\be
{\cal Q}^2=\frac{5}{18}{\ID} +\frac{1}{6}D_{3i}\tau_{i}\ .
\label{qsquare}
\ee
Here $D_{ij}=\frac{1}{2}\
{\rm tr}\left(\tau_{i}A(t)\tau_{j}A^{\dagger}\right)$ denotes the
adjoint representation of the collective rotation which is defined
in eq (\ref{collrot}). The integrals over $\xi^-$ and $\theta$ 
enforce both the constraint 
${\rm cos}\, \theta=({\rm\epsilon_{\rm v}}-Mx)/p$ and the lower bound,  
$p_{\rm min}=|Mx-{\rm\epsilon_{\rm v}|}$ on the $p$ integration. This 
results in the forward moving quark contributions to the transverse 
and longitudinal chiral odd nucleon structure functions
\be
h^{(+)}_T(x)=N_C\frac{2M}{\pi}\int_{p_{\rm min}}^\infty 
pdp\ d\phi\langle \bbox{S}_\perp |\tilde{\psi}_+^\dagger(\bbox{p})
\gamma_\perp\gamma_5 {\cal Q}^2\tilde{\psi}_+(\bbox{p})
|\bbox{S}_\perp\rangle\ \Big|\, _{{\rm cos}\, 
\theta=\frac{{\rm\epsilon}-Mx}{p}}
\ee
and
\be
h^{(+)}_L(x)&=&N_C\frac{2M}{\pi}\int_{p_{\rm min}}^\infty pdp\ d\phi
\langle \bbox{S}_z |\tilde{\psi}_+^\dagger(\bbox{p})
\gamma_0\gamma_5{\cal Q}^2\tilde{\psi}_-(\bbox{p})
\nonumber \\ &&\hspace{3cm}
-\tilde{\psi}_-^\dagger(\bbox{p})
\gamma_0\gamma_5{\cal Q}^2\tilde{\psi}_+(\bbox{p})
|\bbox{S}_z\rangle\ 
\Big|\, _{{\rm cos}\, \theta=\frac{{\rm\epsilon}-Mx}{p}}\ .
\ee
In order to obtain the full structure functions the contribution of 
backward moving quarks $h_{T,L}^{(-)}$ must be considered as well. These 
contributions are easily obtained from $h_{T,L}^{(+)}(-x)$ by 
reversing the appropriate signs in equations (\ref{htpx}) and (\ref{hlpx}). 
The two contributions may be comprised in\footnote{We have used that 
the valence quark level has positive parity, {\it i.e.} under
$\bbox{p}\rightarrow\, -\bbox{p}$ we find
$\tilde{\psi}(-\bbox{p})=\gamma^0\tilde{\psi}(\bbox{p})$.}
\be
h^{(\pm)}_T(x)&=&\pm N_C\frac{M}{\pi}\int_{p_{\rm min}^{\mp}}^\infty 
pdp\ d\phi\langle \bbox{S}_\perp |\tilde{\psi}^\dagger(\bbox{p}_{\mp})
\left(1\mp\alpha_3\right)\gamma_\perp\gamma_5
{\cal Q}^2\tilde{\psi}(\bbox{p}_{\mp})
|\bbox{S}_\perp\rangle\ 
\Big|\, _{{\rm cos}\, \theta^{\mp}_p}
\label{ht11}
\ee
and
\be
h^{(\pm)}_L(x)&=&\pm N_C\frac{M}{\pi}\int_{p_{\rm min}^{\mp}}^\infty pdp\
d\phi \langle \bbox{S}_z |\tilde{\psi}^\dagger(\bbox{p}_{\mp})
\alpha_3\gamma_0\gamma_5{\cal Q}^2\tilde{\psi}(\bbox{p}_{\mp})
|\bbox{S}_z\rangle\ 
\Big|\, _{{\rm cos}\, \theta^{\mp}_p}
\label{hl11}
\ee
where $p_{\rm min}^{\pm}=|Mx\pm{\rm \epsilon_{\rm v}}|$ 
and ${\rm cos}\, \theta^{\pm}=(Mx \pm \epsilon_{\rm v})/p$ and 
$\tilde{\psi}(\bbox{p}_{\pm})
=\tilde{\psi}(p,{\rm cos}\,\theta^{\pm}_{p},\phi)$.
Finally we summarize our results by decomposing the proton structure 
functions into their (iso)scalar and vector components,
\be
h_T(x)&=&h^{I=0}_{T,+}(x)+h^{I=1}_{T,+}(x)\, 
+ \left(h^{I=0}_{T,-}(x)+h^{I=1}_{T,-}(x)\right)\ , \nonumber \\
h_L(x)&=&h^{I=0}_{L,+}(x)+h^{I=1}_{L,+}(x)\, 
+ \left(h^{I=0}_{L,-}(x)+h^{I=1}_{L,-}(x)\right)\ .
\label{hltnjl}
\ee
The isoscalar piece ($I=0$) originates from the unit matrix in the 
decomposition (\ref{qsquare}) while the isovector part ($I=1$) stems from 
the terms involving the collective coordinates. The explicit expressions 
for the structure functions (\ref{hltnjl}) in terms of the static quark
wave functions are computed in Appendix B.

\bigskip
\section{Projection And Evolution}
\bigskip

We consider that our model approximates QCD at a low scale $Q_0^2$. In oder 
to compare the predicted structure functions with data they must be
evolved to a (larger) $Q^2$ commensurate with experimental conditions.
A direct comparison with data gathered at a low scale cannot be made as 
the latter structure functions contain sizable contributions from higher 
twist.  Thus we evolve the chiral odd model structure functions
of the preceding section utilizing the results of perturbative
QCD.

\begin{figure}[ht]
\centerline{
\epsfig{figure=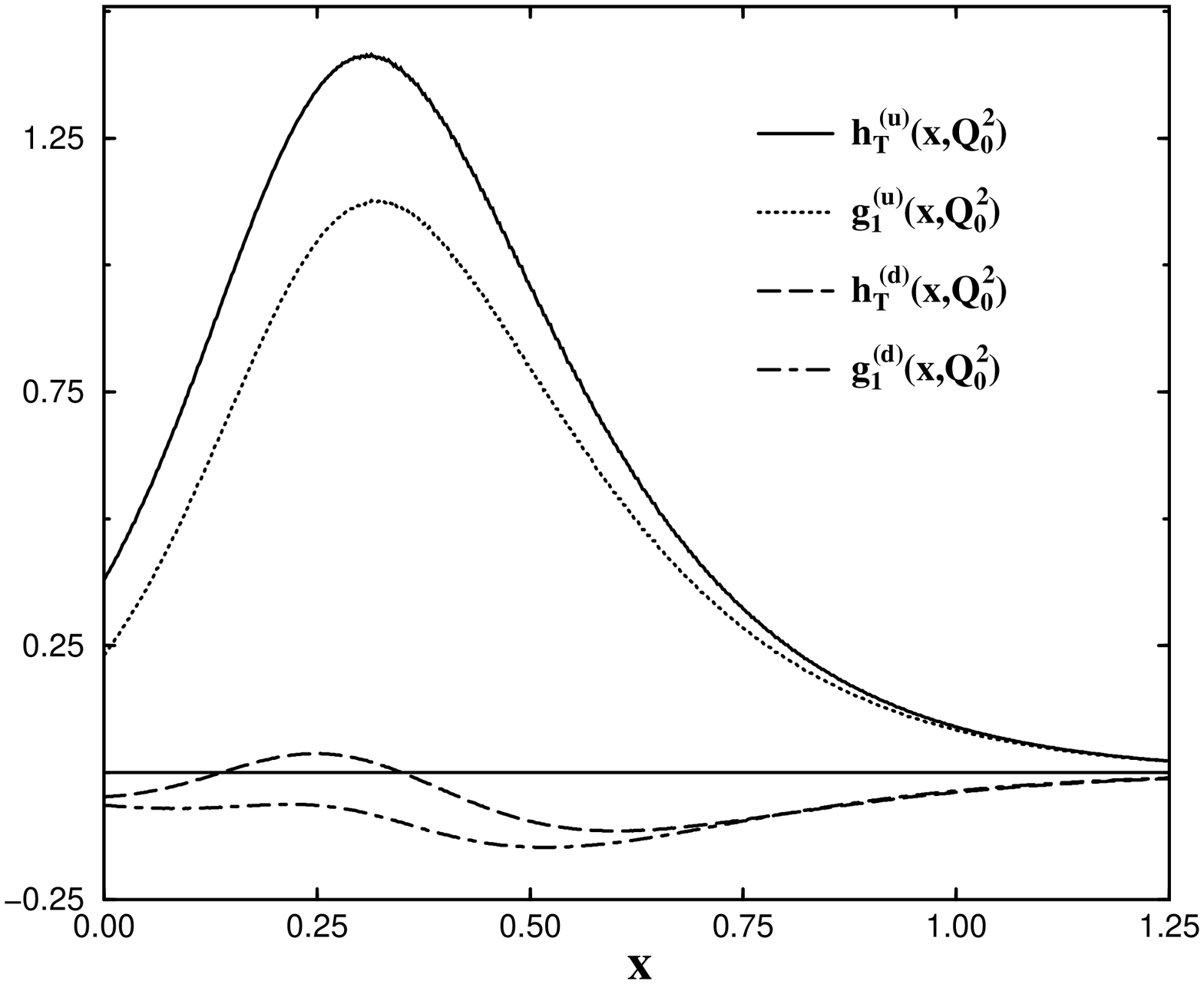,height=8.5cm,width=8.0cm,angle=270}
\hspace{-0.5cm}
\epsfig{figure=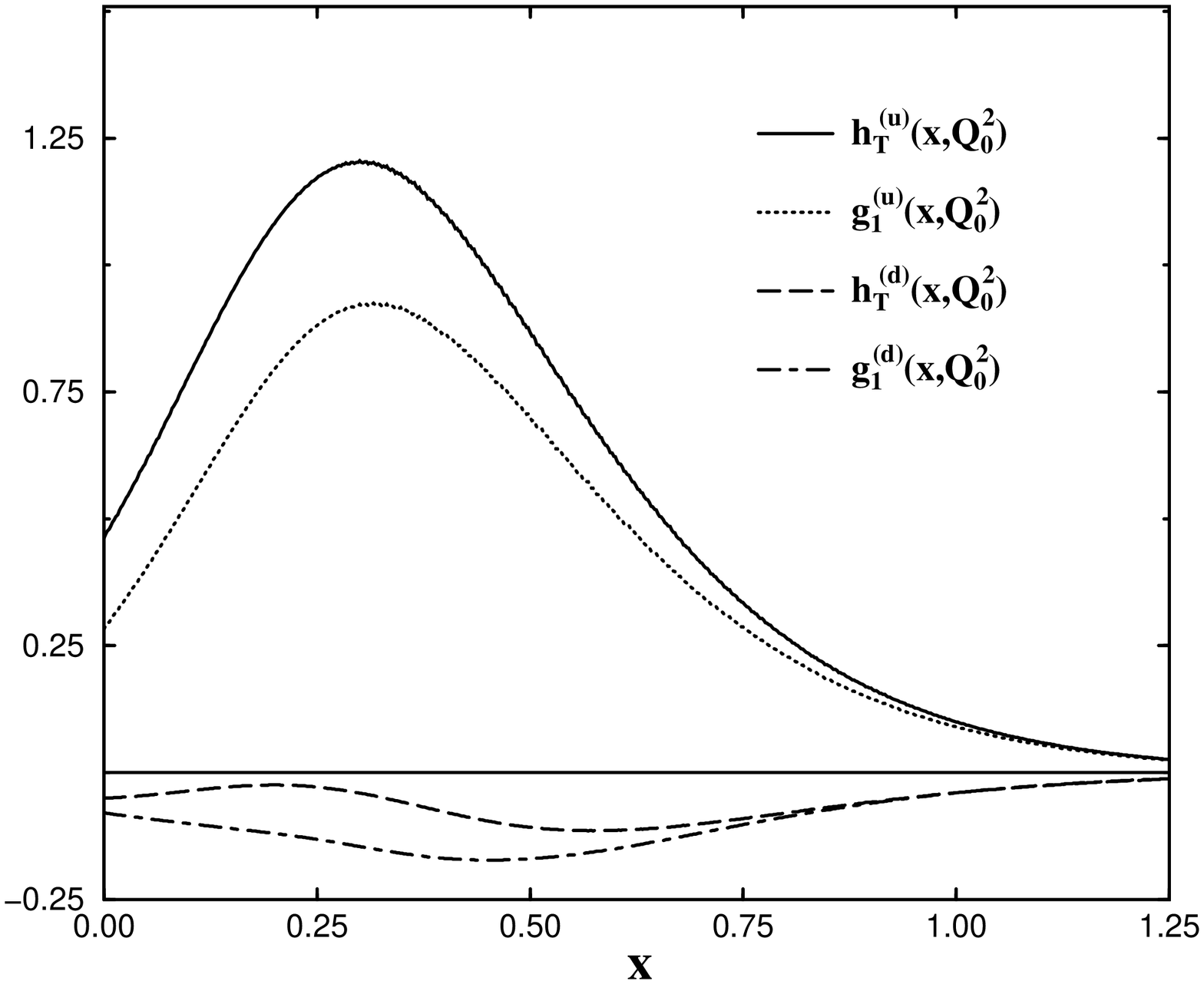,height=8.5cm,width=8.0cm,angle=270}}
\caption{\label{fig_htud}
The valence quark approximation of the transverse chiral--odd nucleon 
distribution function as a function of Bjorken--$x$ for the up and down 
quark flavor content in the rest frame. For comparison also the model 
calculation \protect\cite{We97} for the twist--2 polarized structure 
function $g_1(x,Q_0^2)$ is shown for the respective flavor channels.
Two values of the constituent quark mass are considered:
$m=400 {\rm MeV}$ (left panel) and $m=450 {\rm MeV}$ (right panel).}
\end{figure}

In the soliton approach the baryon states are built from localized 
field configurations. In fact, these states do not carry good 
four--momentum. Therefore the calculated structure functions 
({\it cf.} Figs. \ref{fig_htud}, \ref{fig_htnp} and \ref{fig_hlnp}) do 
not vanish exactly for $x>1$ although the contributions for $x>1$ 
are very small. 

The calculation of nucleon 
structure functions in the Bjorken limit, however, singles out 
the null plane, $\xi^+=0$. This condition can be satisfied upon 
transformation to the infinite momentum frame (IMF) even for models 
where the nucleon emerges as a (static) localized object \cite{Hu77}.  
For the quark soliton model under consideration this transformation 
corresponds to performing a boost in the space of the collective 
coordinate $\bbox{x}_0$, {\it cf.} eq (\ref{cht}). Upon this boost 
to the IMF we have observed \cite{Ga97} that the common problem of 
improper support for the structure functions, {\it i.e.} non--vanishing 
structure functions for $x>1$, is cured along the line suggested by 
Jaffe \cite{Ja80} some time ago. The reason simply is that the Lorentz 
contraction associated with the boost to the IMF maps the infinite line 
exactly onto the interval $x\in [0,1[$. In addition we have observed that 
this Lorentz contraction effects the structure functions also at small 
and moderate $x$.  Incorporating these results for the general set 
of leading twist structure functions within the NJL--chiral soliton model
yields the following form for the forward and backward
moving  intermediate quark
state contributions to the chiral odd transverse
spin structure function, $h^{(\pm)}_T\left(x,Q^2\right)$,
\be
\hspace{-0.3cm}
h^{(\pm)}_T(x)&=&\pm N_C\frac{M}{\pi(1-x)}
\int_{p_{\rm min}}^\infty \hspace{-0.2cm} pdp d\varphi \
\nonumber \\ && \hspace{1.0cm}
\times\langle N |\tilde{\psi}^\dagger (\bbox{p}_{\mp})
\left(1\mp\alpha_3\right)\gamma_{\perp}\gamma_5{\cal Q}^2
\tilde{\psi}(\bbox{p}_{\mp})|N\rangle
\Big|_{{\rm cos}\theta=-
{\textstyle \frac{M\ {\rm ln}(1-x)\pm\epsilon_{\rm v}}{p}}} \ .
\label{htp}
\ee
In general the resulting relation between structure functions 
in the IMF and the rest frame (RF) reads
\be
f_{\rm IMF}(x)=\frac{\Theta(1-x)}{1-x} f_{\rm RF}
\Big(-{\rm ln}(1-x)\Big)\ .
\label{fboost}
\ee
Of course, in the context of the chiral odd structure functions 
$f_{\rm RF}$  is to be identified with the expressions in 
eqs (\ref{ht11},\ref{hl11},\ref{hltnjl}). As will be recognized 
shortly the solution to
the proper support problem is essential in order to 
apply the evolution program of perturbative QCD.
\begin{figure}[ht]
\centerline{
\epsfig{figure=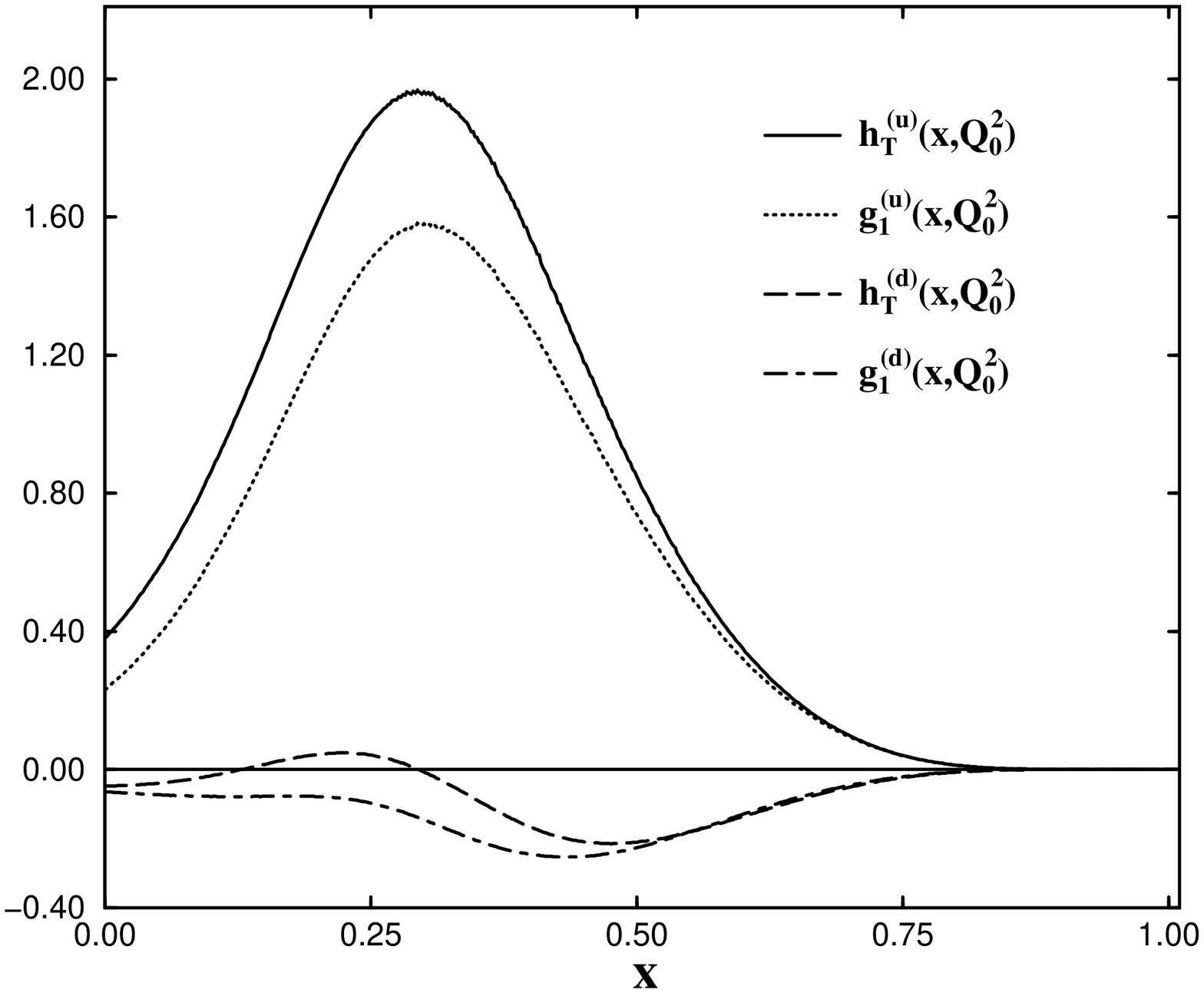,height=8.5cm,width=8.0cm,angle=270}
\hspace{-0.5cm}
\epsfig{figure=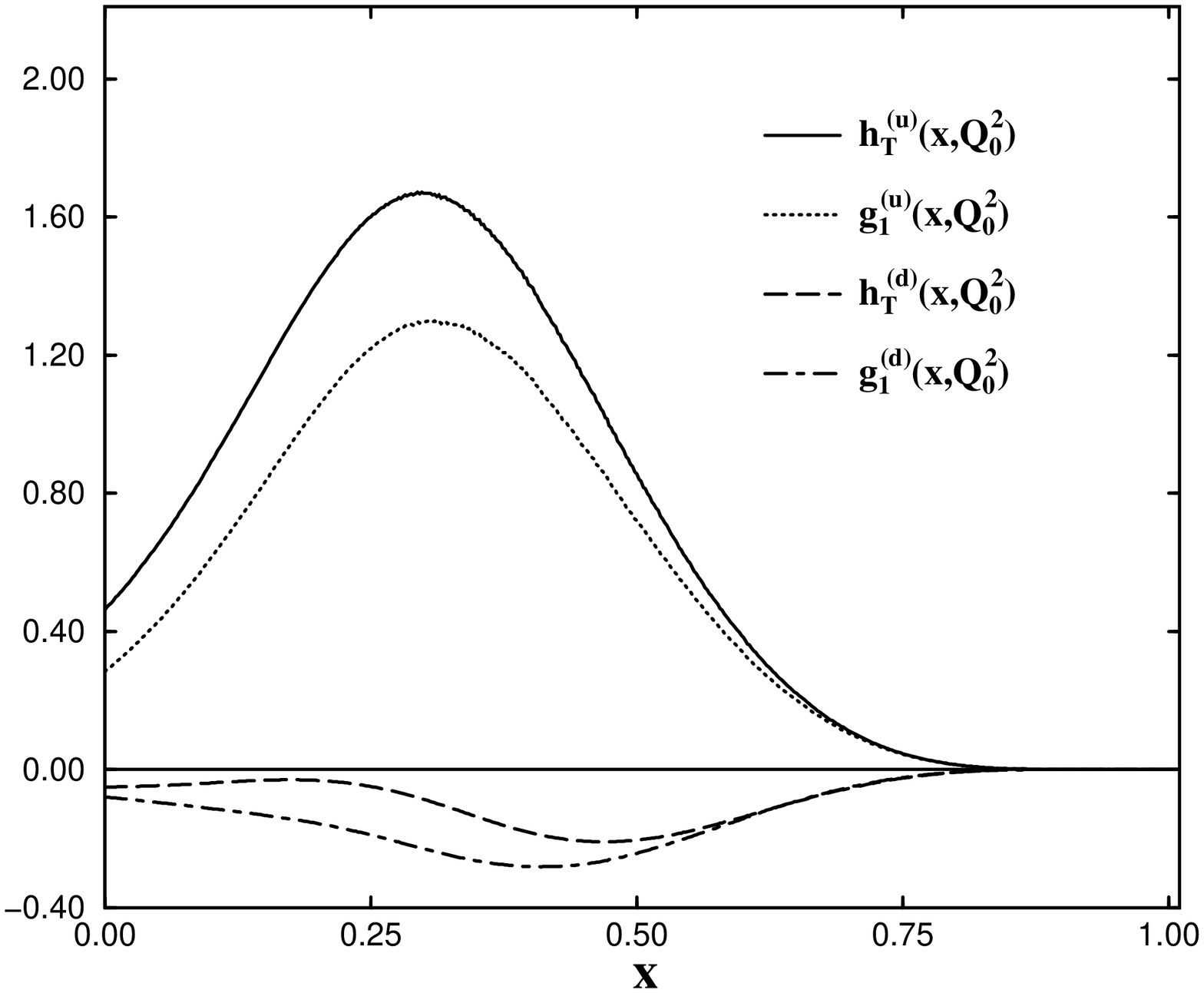,height=8.5cm,width=8.0cm,angle=270}}
\caption{\label{fig_htudpr}
Same as figure \protect\ref{fig_htud} 
in the IMF (\protect\ref{fboost}).}
\end{figure}
The chiral odd and polarized structure functions resulting from this
transformation are shown in figure \ref{fig_htudpr}.

In order to include the logarithmic corrections to the 
twist--2 pieces of the chiral odd structure functions we 
apply the well--established GLAP procedure \cite{Gr72}.
For the transverse component $h_T(x,Q^2)$ this is 
straightforward as it is pure twist--2. For the longitudinal
piece $h_L(x,Q^2)$ one first has to extract the twist--2
component through $h_T(x,Q^2)$ namely,
$h_L^{(2)}(x,Q^2)=2x\, \int_{x}^1\ dy\,h_T(y,Q^2)/y^2$.

We simultaneously denote by $h^{(2)}$ the twist--2 parts of $h_T$ 
and $h_L$. To leading order (in $\alpha_{QCD}(Q^2)$) the variations
of the structure functions from a change $\delta t$ of the 
momentum scale is given by
\be
h^{(2)}(x,t+\delta{t})=h^{(2)}(x,t)\ 
+ \frac{dh^{(2)}(x,t)}{dt} \ \delta t\ ,
\label{h2var}
\ee
where $t={\rm log}\left(Q^2/\Lambda_{QCD}^2\right)$. The variation
(\ref{h2var}) is essentially due to the emission and absorption of 
soft gluons. The explicit expression for the evolution differential 
equation is given by the convolution integral,
\be
\frac{d\, h^{(2)}(x,t)}{dt} =\frac{\alpha_{QCD}(t)}{2\pi}
C_{R}(F)\int^1_{x}\ \frac{dy}{y}P_{qq}^h\left(y\right)
h^{(2)}\left(\frac{x}{y},t\right)
\label{convl}
\ee
where the leading order splitting function \cite{Ar90,Ba97} is
given by,
\be 
P_{qq}^{h}\left(z\right)=
\frac{4}{3}\left[\frac{2}{\left(1-z\right)_+}-2
+\frac{3}{2}\ \delta(z-1)\right]
\ee
and $C_R(f)=\left(n_f^2-1\right)/2n_f$ for $n_f$ active flavors,
$\alpha_{QCD}(t)=4\pi/\left[b_0\log\left(Q^2/ \Lambda^2\right)\right]$
and $b_0=(11N_C-2n_f)/3$.
Employing the ``+" prescription yields for three light flavors and
$N_C=3$
\be
\frac{d h^{(2)}(x,t)}{dt}&=&\frac{\alpha_{QCD}(t)}{2\pi}
\left\{\ \left(2 + \frac{8}{3}\log(1-x)\right)h^{(2)}(x,t)
\right.
\nonumber \\*&& \hspace{-0.7cm} 
\left. 
+\, \frac{8}{3}\int^{1}_{x}\ \frac{dy}{y}\left[
\frac{1}{1-y}\left(h^{(2)}(\frac{x}{y},t)-yh^{(2)}(x,t)\right)
- h^{(2)}(\frac{x}{y},t)\right]\right\}\ .
\label{evhtw2}
\ee
As indicated above, the structure functions must vanish at the boundary 
$x=1$ in order to cancel the divergence of the logarithm in eq 
(\ref{evhtw2}) and thus for the GLAP procedure to be applicable. This 
makes the projection of the rest frame structure functions mandatory.
The variation of the structure functions for finite intervals 
in $t$ is straightforwardly obtained by iteration of these 
equations, {\it i.e.} as a solution to the differential 
equation (\ref{evhtw2}). As discussed previously the initial value 
for integrating the differential equation is given by the scale 
$Q_0^2$ at which the model is defined. It should be emphasized that 
this scale essentially is a new parameter of the model. For a given 
constituent quark mass $m$ we adjust $Q_0^2$ to maximize the 
agreement of the predictions with the experimental data on 
previously \cite{We96a} calculated unpolarized structure functions for 
electron--nucleon DIS: $F_2^{ep}-F_2^{en}$. For the constituent 
quark mass $m=400{\rm MeV}$ we have obtained $Q_0^2\approx0.4{\rm GeV}^2$.
Note that this value of $Q_0^2$ is indeed (as it should) smaller than 
the ultraviolet cut--off of the underlying NJL soliton model as 
$\Lambda^2\approx 0.56{\rm GeV}^2$. The latter quantity indicates the range 
of validity of the model. In figure \ref{fig_ht2p}a we compare the un--evolved, 
projected, proton structure function $h_T^{p}\left(x,Q_0^2\right)$ with 
the one evolved from $Q_0^2=0.4{\rm GeV}^2$ to $Q^2=4.0{\rm GeV}^2$. As 
expected the evolution pronounces the structure function at low $x$. 

This change towards small $x$ is a generic feature of the projection 
and evolution process and presumably not very sensitive to the 
prescription applied here. In particular, choosing a projection 
technique \cite{Tr97} alternative to (\ref{fboost}) may easily be 
compensated by an appropriate variation of the scale $Q_0^2$. In 
figure \ref{fig_ht2p}b the same calculation for $h_L^{(2)}(x,Q^2)$ is 
presented.

In the evolution of the twist--2 pieces we have restricted ourselves
to the leading order in $\alpha_s$ because for the twist--3 piece of
$h_L$, the necessary ingredients are not known in next--to--leading 
order. Even the leading order evolution is only known in the large 
$N_C$ limit. It should be noted that such an approach seems 
particularly suited for soliton models which equally utilize large 
$N_C$ arguments. As pointed out by Balitskii et al. \cite{Bal96} the 
admixture of independent quark and quark--gluon operators contributing 
to the twist--3 portion ${\overline{h}}_L(x,Q^2)$ grows with $n$ 
where $n$ refers to the $n^{\rm th}$ moment,
${\cal M}_n\left[ \overline{h}_L(Q^2)\right]$ of $h_L(x,Q^2)$.
However, much like the case with
the spin--polarized structure function, $g_2(x,Q^2)$ \cite{Ali91}
in the $N_C\rightarrow \infty$ limit the quark operators of 
twist--3 decouple from the quark--gluon operators of the same twist.
Then the anomalous dimensions $\gamma_n$ which govern the 
logarithmic $Q^2$ dependence of ${\cal M}_n$ can be computed. Once the 
$\gamma_n$'s are known an evolution kernel can be constructed that 
``propagates'' the the twist--3 part $\overline{h}(x,Q^2)$ in momentum
\be
\overline{h}_L(x,Q^2)&=&\int_x^1 \frac{dy}{y} b(x,y;Q^2,Q_0^2)
\overline{h}_L(y,Q_0^2)\ .
\label{evkern}
\ee
We relegate the detailed discussion of the kernel $b(x,y;Q^2,Q_0^2)$,
which is obtained by inverting the $Q^2$ dependence of ${\cal M}_n$,
to appendix C. In figure \ref{fig_h2bllp}a we show the evolution of 
$\overline{h}_L(x)$. Again we used $Q_0^2=0.4{\rm GeV}^2$ and 
$Q^2=4.0{\rm GeV}^2$.

As discussed in ref \cite{Bal96} the merit of this 
approach is that to leading order in $N_C$ the knowledge of 
$h_L(x,Q^2)$ at one scale is sufficient to predict it at any arbitrary 
scale, which is not the case at finite $N_C$.\footnote{As noted in 
\cite{Bal96}, next to leading order corrections are estimated to go 
like $O\left(1/N^2_c\times{\rm ln}(n)/\, n\right)$ at large $n$.}
Thus $h_L(x,Q^2)$ obeys a generalized GLAP evolution equation. 
This finally enables us (in much the same manner as was the case 
for $g_2(x,Q^2)$ in \cite{We97}) to compute the longitudinal chiral odd 
structure function $h_L(x,Q^2)$ by combining the separately evolved 
twist--2 and twist--3 components together. The result for 
$Q_0^2=0.4{\rm GeV}^2$ and $Q^2=4.0{\rm GeV}^2$ is shown in figure
\ref{fig_h2bllp}b. We recall that the only ingredients have been the leading 
twist pieces of the chiral odd structure functions at the model 
scale $Q_0$.\footnote{A feature of $h_L(x)$ compared with $g_2(x)$ 
is that as $h_L(x)$ does not mix with gluon distributions 
owing to its chiral-odd nature and its $Q^2$ evolution is given by 
(\ref{mom}), (\ref{adm}) even for the flavor singlet piece.} 

\bigskip
\section{Discussion of the Numerical Results}
\bigskip
In this section we discuss the results of the chiral-odd structure
functions calculated from eqs (\ref{hT0})--(\ref{hL1}) for constituent
quark masses $m=400 {\rm MeV}$ and $m=450 {\rm MeV}$. In figure
\ref{fig_htud} we have shown the up and down quark contributions
to the transverse chiral odd structure function of the proton. Figure
\ref{fig_htudpr} displays them boosted to the IMF.  We observe
that these structure functions are always smaller (in magnitude) than
the twist--2 polarized structure function $g_1$ with the same flavor
content. This relation is also known from the bag model \cite{Ja92}.
Similar to the confinement model calculation of Barone {\it et al.}
\cite{Ba97} we find that $h_T^{(d)}(x)$ is negative at small $x$. In
contrast to $g_1^{(d)}(x)$, however, it might change sign although
the positive contribution appears to be small and diminishing with
increasing constituent quark mass.

As already indicated in the introduction the DIS processes which are 
sensitive to these distributions will provide access to the charge 
weighted combinations thereof. We will hence concentrate on this flavor 
content. In any event, as we will be discussing both, the proton and 
the neutron chiral odd distributions, other flavor combinations can 
straightforwardly be extracted by disentangling the isoscalar 
and isovector pieces in eq (\ref{qsquare}). In 
connection with the chiral--odd transverse nucleon structure function 
we also calculate its zeroth moment which is referred to as the isoscalar 
and isovector nucleon tensor charges \cite{Ja92},
\be
\Gamma^S_{T}(Q^2) &=& \frac{18}{5} \int_0^1\, 
\left[ dx\ h_T^p\left(x,Q^2\right)\
+ h_T^n\left(x,Q^2\right)\right]
\label{gtens} \\
\Gamma^V_{T}(Q^2) &=& 6 \int_0^1\, \left[ dx\ h_T^p\left(x,Q^2\right)\ 
- h_T^n\left(x,Q^2\right)\right] 
\label{gtenv}
\ee
at both the low scale, $Q_0^2=0.4 {\rm GeV}^2$ and a scale commensurate
with experiment, $Q^2= 4 {\rm GeV}^2$. Of course, for the neutron we 
have to reverse the signs of the isovector pieces in eq (\ref{hltnjl}).
In eqs (\ref{gtens}) and (\ref{gtenv}) the normalization factors are 
due to the separation into isosinglet and isovector contributions, 
{\it cf.} eq (\ref{qsquare}). Note that due to 
$\int_0^1 dz P_{qq}^h(z)\ne0$ the tensor charge is not protected against 
logarithmic corrections. Our results for the valence quark approximation 
are summarized in Table 1. For completeness we also add the vacuum 
contribution to the tensor charges at the model scale $Q_0^2$. Their 
analytic expressions are given in appendix D. Obviously this 
vacuum contribution is negligibly small. This is a strong
justification of the valence quark approximation to the chiral 
odd structure functions. 
\begin{table}[ht]
\caption{\label{tab_1}
Nucleon tensor charges calculated from eqs (\ref{gtens}) and
(\ref{gtenv}) as a function of the constituent quark mass $m$ in the
NJL chiral--soliton model. The momentum scales are $Q_0^2=0.4{\rm GeV}^2$
and $Q^2=4.0{\rm GeV}^2$. The numbers in parenthesis in the respective
upper rows include the negligible contribution from the polarized quark
vacuum. We compare with results from the Lattice \protect\cite{Ao97},
QCD sum rules \protect\cite{He95}, the constituent quark model with
Goldstone boson effects \protect\cite{Su97} and a quark soliton model 
calculation \protect\cite{Ki96} including multiplicative $1/N_C$ corrections 
violating PCAC in the similar case of the axial vector current 
\protect\cite{Al93}. Finally the predictions from the confinement model 
of ref \protect\cite{Ba97} with the associated momentum scales 
(in ${\rm GeV}^2$) are shown.}
~ \vskip0.1cm
\centerline{
\renewcommand{\arraystretch}{1.5}
\begin{tabular}{c|lll|llll|ll}
$m$ ({\rm MeV})  &~~~350 &~~~400  &~~~450 
& Lat. & ~SR & ~CQ & ~QS & ~$Q^2$ & CM
\\ \hline
$\Gamma^S_T(Q_0^2)$ & 0.80 (0.82)
& 0.72 (0.76) & 0.67 (0.72)
& 0.61 & 0.61 & 1.31 & 0.69 & 0.16 & 0.90 \\
$\Gamma^S_T(Q^2) $ & 0.73  & 0.65 & 0.61 
&\multicolumn{4}{c|}{no scale attributed} 
&25.0 & 0.72\\
\hline
$\Gamma^V_T(Q_0^2)$ & 0.88 (0.89) 
& 0.86 (0.87) & 0.86 (0.85) 
& 1.07 & 1.37 & 1.07 & 1.45 & 0.16 & 1.53 \\
$\Gamma^V_T(Q^2) $ & 0.80  & 0.78 & 0.77 
&\multicolumn{4}{c|}{no scale attributed}
&25.0 & 1.22 \\
\end{tabular}}
\renewcommand{\arraystretch}{1.0}
\end{table}
A further justification comes from a recent
study of the Gottfried sum rule within the same model \cite{Wa98}.
Also in that case the contribution of the distorted quark vacuum
to the relevant structure function turned out to be negligibly
small.

Besides justifying the valence quark approximation for the chiral 
odd distributions table \ref{tab_1} contains the comparison to other 
model calculations of the nucleon tensor charges. We note that in obtaining 
the isovector tensor charge $\Gamma_T^V$ we have omitted contributions 
which are suppressed by $1/N_C$ ({\it cf.} appendix D). These contributions
arise when one adopts a non--symmetric ordering of the operators in 
the space of the collective operators \cite{Ki96}. The main reason for 
taking the symmetric ordering is that in the case of the isovector axial 
charge, $g_A$, any non--symmetric ordering of the collective operators 
leads to a sizable violation of PCAC unless the meson profile is not 
modified \cite{Al93}. These multiplicative $1/N_C$ corrections \cite{Da94} 
may be the reason why our predictions for $\Gamma_T^V$ are somewhat lower 
than those of other models. In the case of the flavor singlet component, 
which does not have such corrections, our results compare nicely with 
other model calculations except for the constituent quark model of 
ref \cite{Su97}.

\begin{figure}[ht]
\centerline{
\epsfig{figure=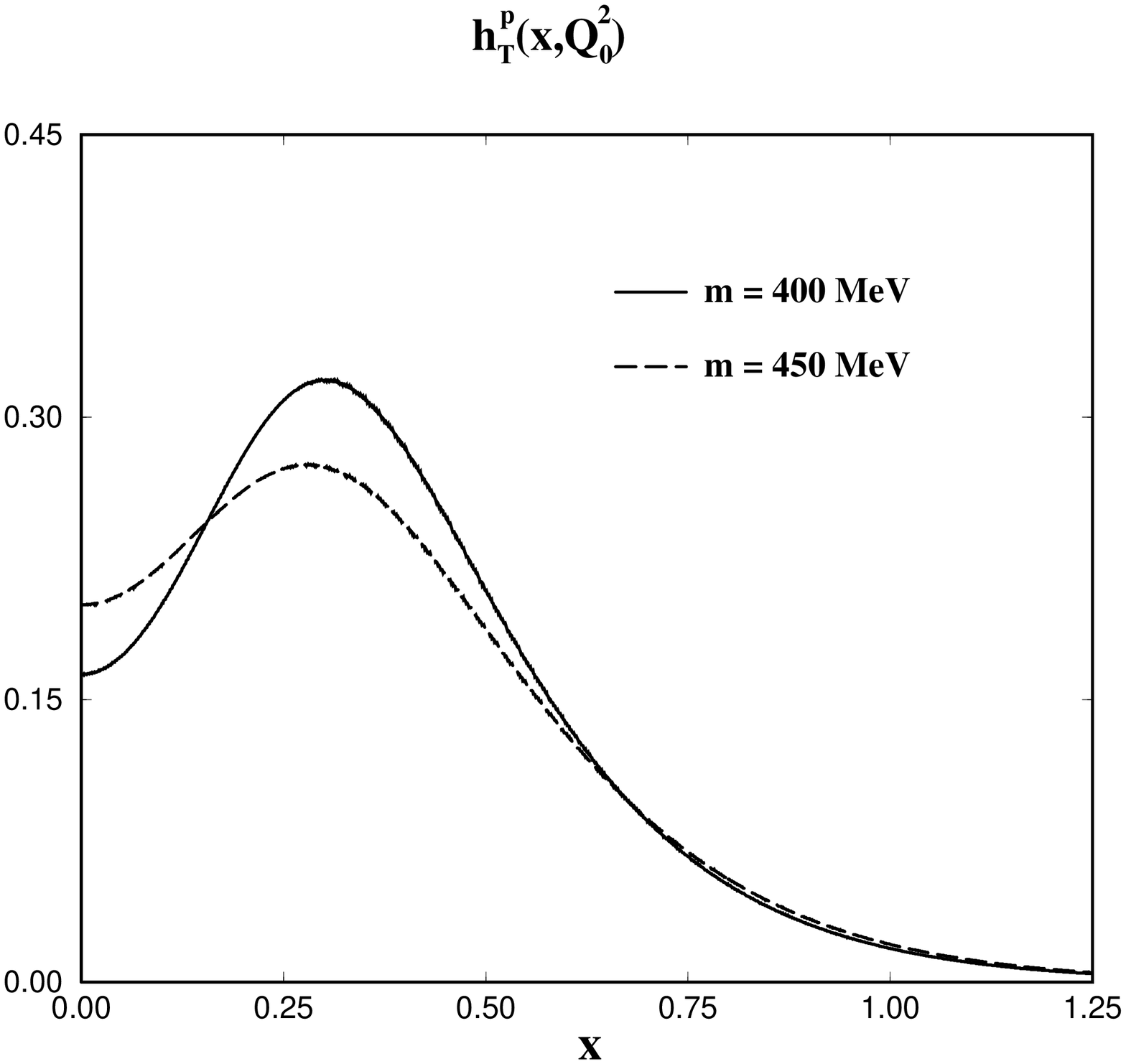,height=8.5cm,width=8.0cm,angle=270}
\hspace{-0.5cm}
\epsfig{figure=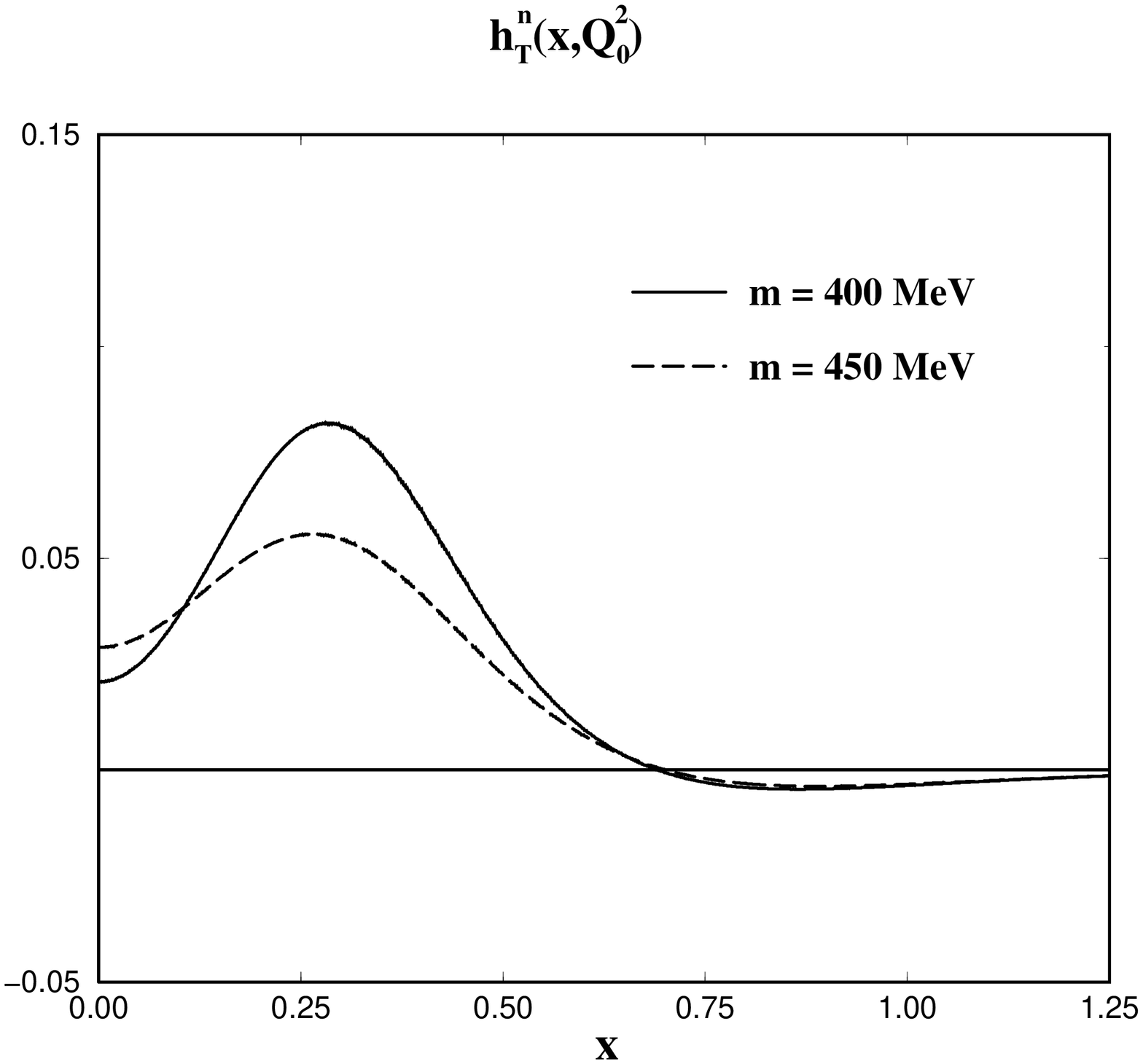,height=8.5cm,width=8.0cm,angle=270}}
\caption{\label{fig_htnp}
The valence quark approximation of the chiral--odd
nucleon structure functions as a function of Bjorken--$x$.
Left panel: $h_{T}^{p}\left(x ,Q_0^2\right)$ for constituent
quark masses $m=400 {\rm MeV}$ (solid line) and
$m=450 {\rm MeV}$ (long--dashed line).
Right panel: $h_{T}^{n}\left(x,Q_0^2\right)$.}
\end{figure}

\begin{figure}[ht]
\centerline{
\epsfig{figure=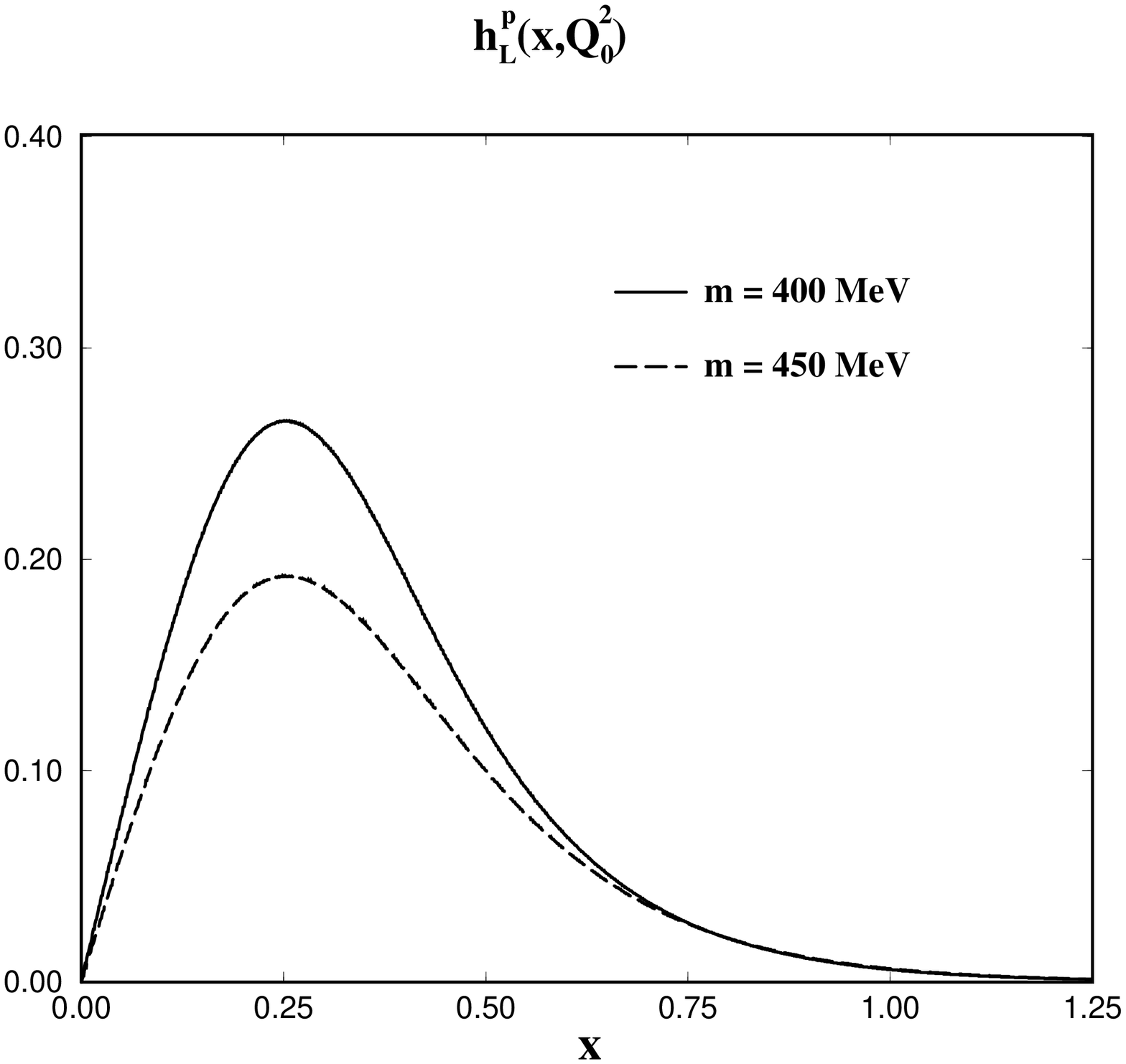,height=8.5cm,width=8.0cm,angle=270}
\hspace{-0.5cm}
\epsfig{figure=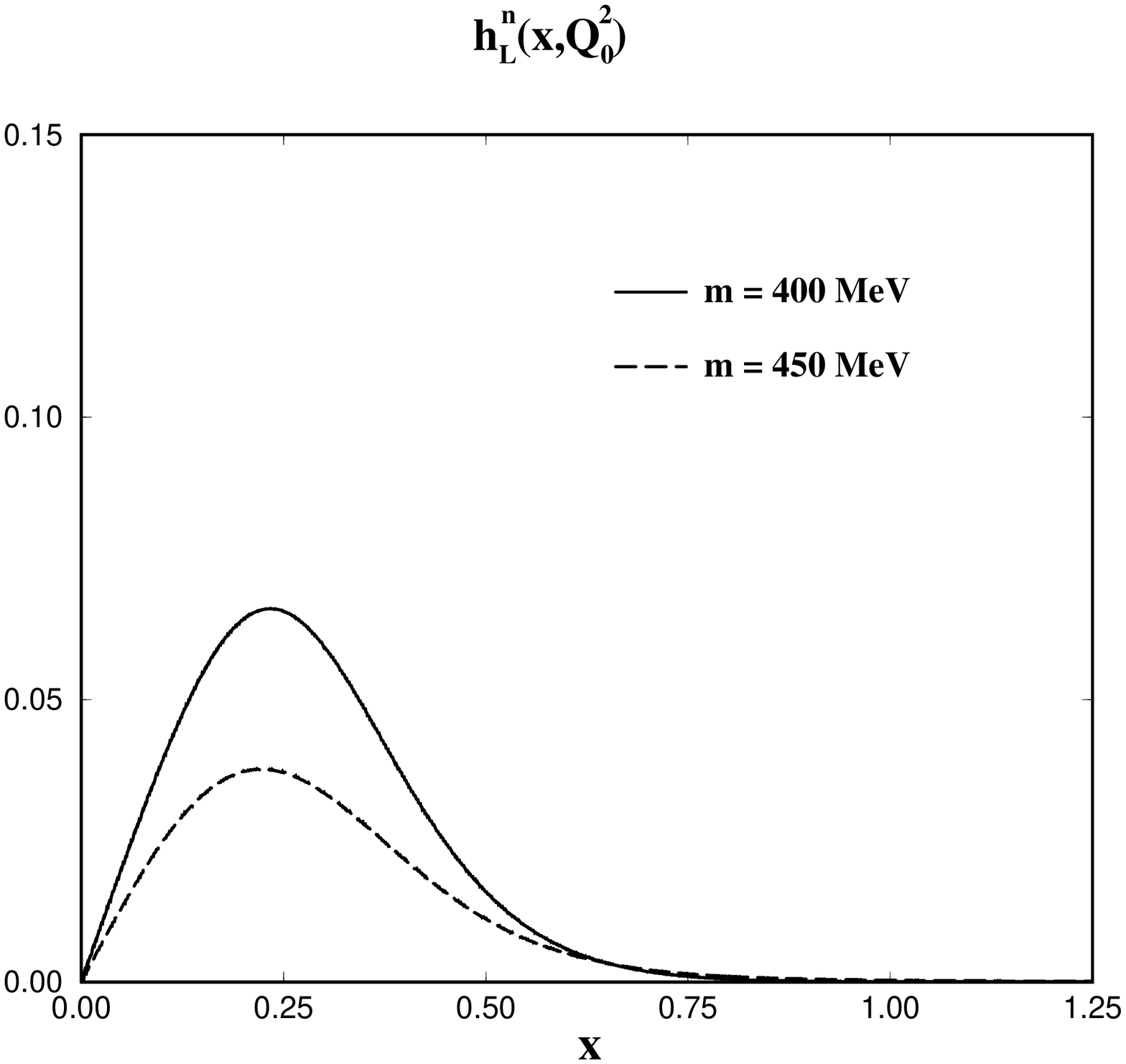,height=8.5cm,width=8.0cm,angle=270}}
\caption{\label{fig_hlnp}
The valence quark approximation of the chiral--odd
nucleon structure functions as a function of Bjorken--$x$.
Left panel: $h_{L}^{p}\left(x ,Q_0^2\right)$ for constituent
quark masses $m=400 {\rm MeV}$ (solid line)
and $m=450 {\rm MeV}$ (long--dashed line).
Right panel: $h_{L}^{n}\left(x,Q_0^2\right)$.}
\end{figure}

In figure \ref{fig_htnp} we display the transverse chiral odd proton 
$h_T^{p}\left(x,Q_0^2\right)$ and neutron 
$h_T^{n}\left(x,Q_0^2\right)$ structure functions at the low momentum 
scale $Q_0^2$, while in figure \ref{fig_hlnp} we do the same for the 
corresponding chiral odd longitudinal structure functions 
$h_L^{p}\left(x,Q_0^2\right)$
and $h_L^{n}\left(x,Q_0^2\right)$.
We observe that the structure 
functions $h_{T}^{N}(x,Q_0^{2})$ and $h_{L}^{N}(x,Q_0^{2})$ are 
reasonably localized in the interval $0\le x\le1$. In particular, this
is the case for the chiral odd structure functions of the neutron. 
Nevertheless a projection as in eq (\ref{fboost}) is required to 
implement Lorentz covariance. In addition the computed structure functions 
exhibit a pronounced maximum at $x\approx0.3$ which is smeared out when the 
constituent quark mass $m$ increases. This can be understood as follows:
In our chiral soliton model the constituent mass serves as a coupling
constant of the quarks to the chiral field (see eqs (\ref{bosact})
and (\ref{hamil})). The valence quark level becomes more strongly bound 
as the constituent quark mass increases. Hence the lower components of 
the valence quark wave--function increase with $m$ and relativistic 
effects become more important. This effect results in the above 
mentioned broadening of the maximum.

\begin{figure}[ht]
\centerline{
\epsfig{figure=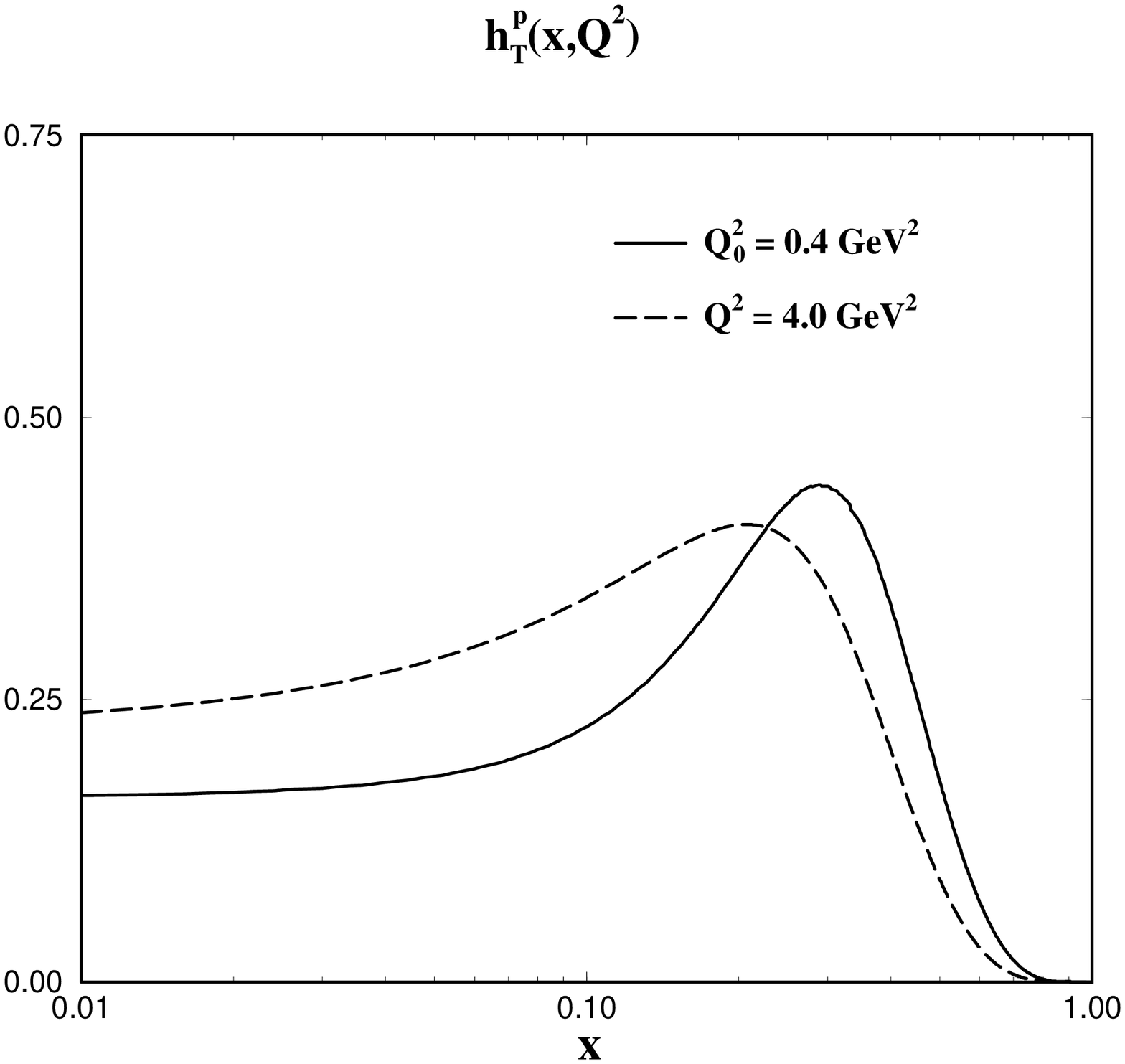,height=8.5cm,width=8.0cm,angle=270}
\hspace{-0.5cm}
\epsfig{figure=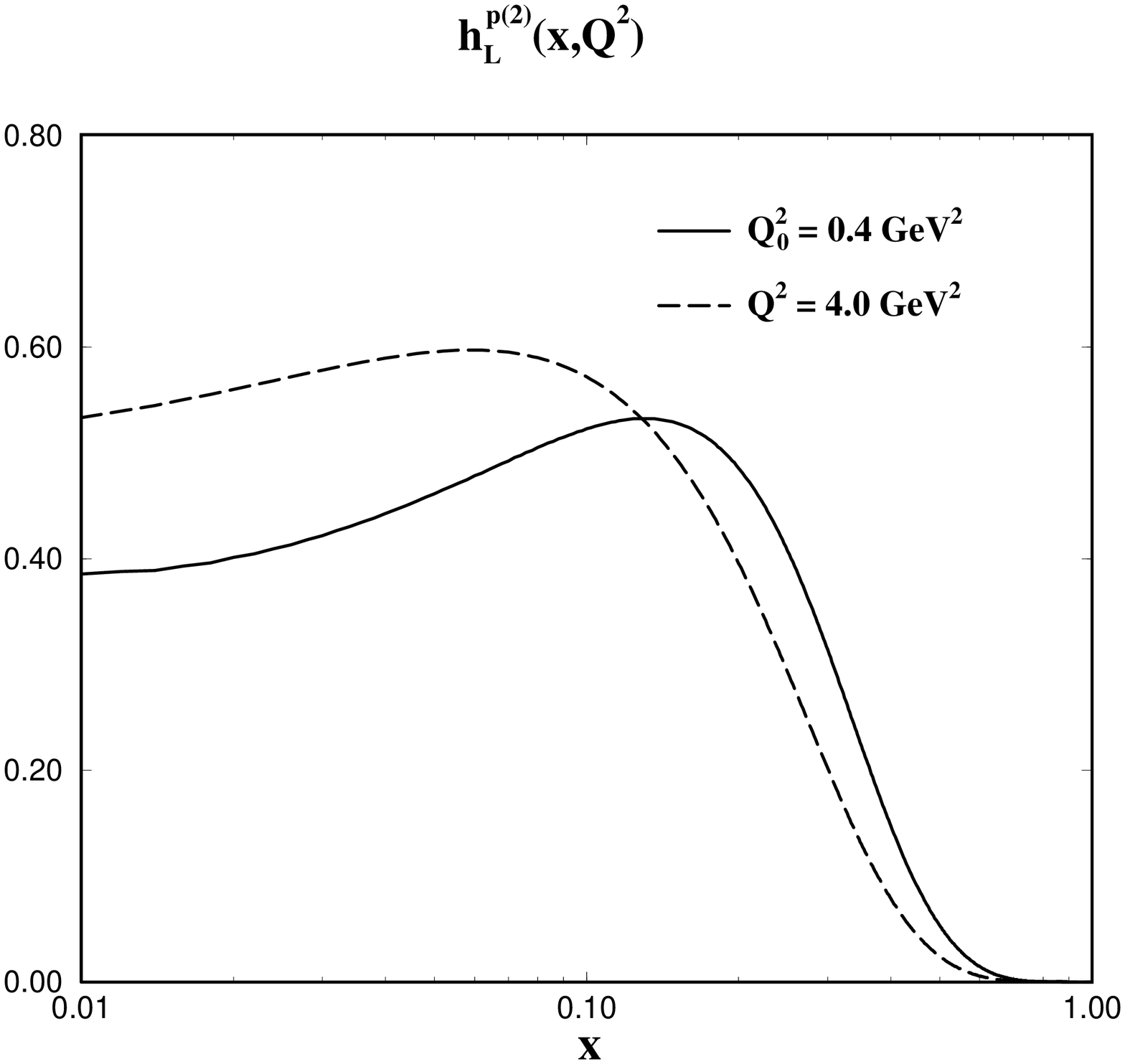,height=8.5cm,width=8.0cm,angle=270}}
\caption{\label{fig_ht2p}
Left panel: The evolution of $h_{T}^{p}\left(x ,Q^{2}\right)$
from $Q^2_0=0.4 {\rm GeV}^2$ (solid line) to $Q^2=4 {\rm GeV}^2$ 
(long--dashed line) for the constituent quark mass $m=400 {\rm MeV}$.
Right panel: The evolution of the twist--2 contribution to the 
longitudinal chiral odd structure function,
$h_{L}^{p(2)}\left(x ,Q^{2}\right)$
from $Q^2_0=0.4 {\rm GeV}^2$ (solid line) to
$Q^2=4 {\rm GeV}^2$ (long--dashed line) for $m=400 {\rm MeV}$.}
\end{figure}

\begin{figure}[ht]
\centerline{
\epsfig{figure=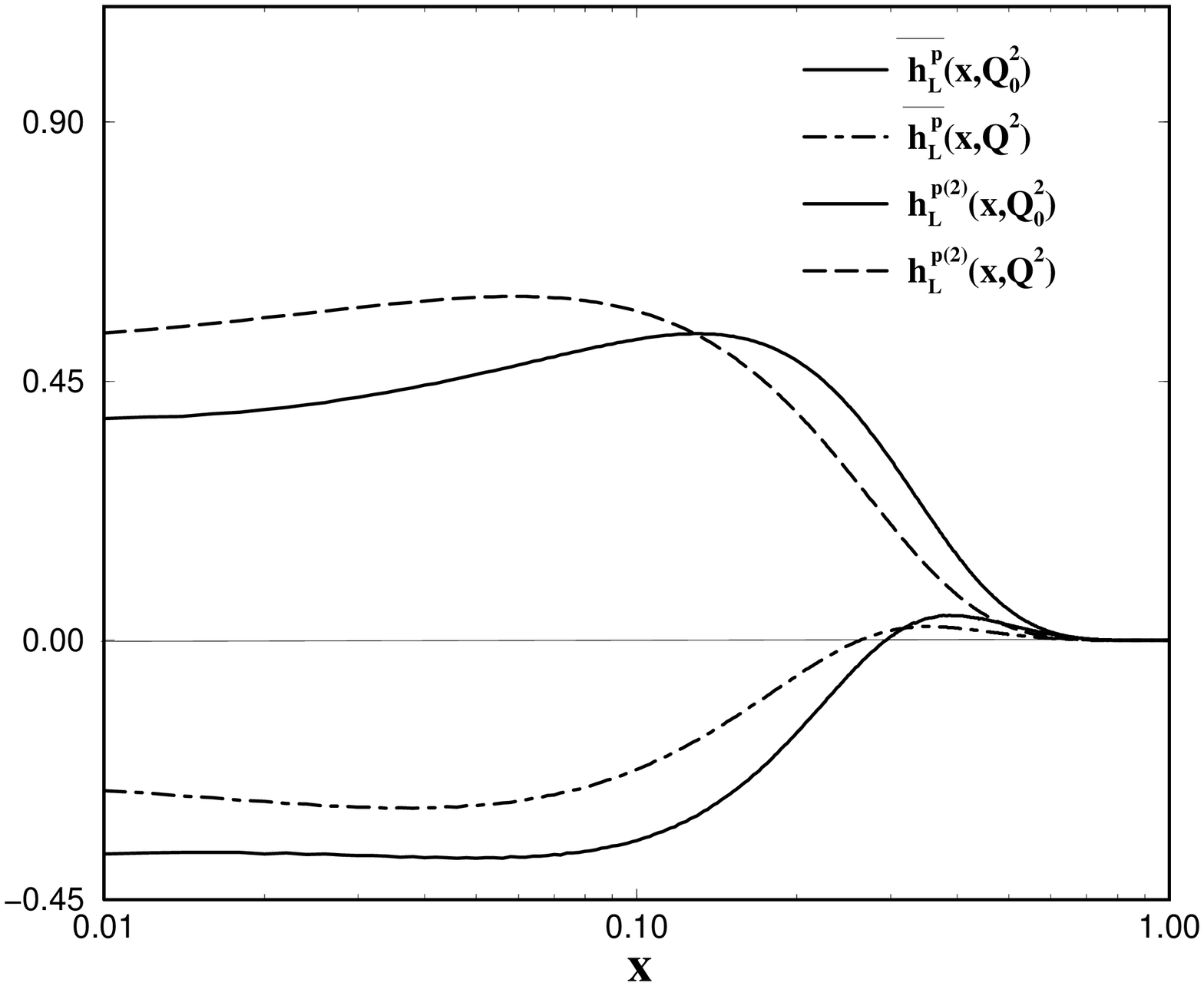,height=8.5cm,width=8.0cm,angle=270}
\hspace{-0.5cm}
\epsfig{figure=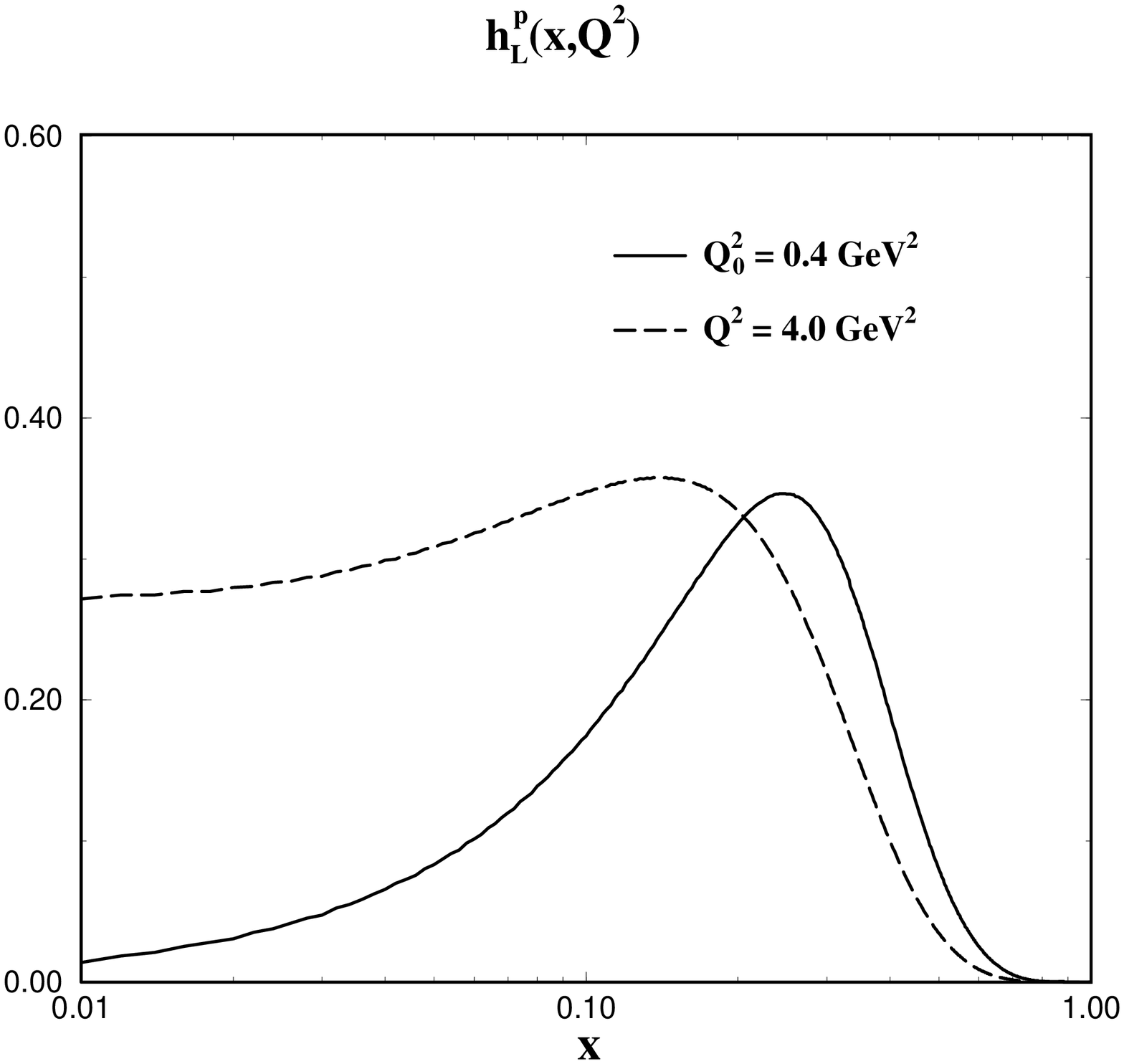,height=8.5cm,width=8.0cm,angle=270}}
\caption{\label{fig_h2bllp}
Left panel (\protect\ref{fig_h2bllp}a): 
The evolution of the twist--3 contribution to the longitudinal 
chiral odd structure function, $\overline{h}_L^p(x,Q^2)$
along with the corresponding twist--2 piece,
$h_{L}^{p(2)}\left(x ,Q^{2}\right)$.
Right panel (\protect\ref{fig_h2bllp}b): The evolution
of $h_{L}^{p}\left(x ,Q^{2}\right)=h_{L}^{p(2)}\left(x ,Q^{2}\right)
+\overline{h}_L^p(x,Q^2)$ from $Q^2_0=0.4 {\rm GeV}^2$ (solid line) to
$Q^2=4 {\rm GeV}^2$ (long--dashed line) for the constituent
quark mass $m=400 {\rm MeV}$.}
\end{figure}

As discussed above a sensible comparison with (eventually available)
data requires either to evolve the model results upward according to
the QCD renormalization group equations or to compare the model 
results with a low momentum scale parameterization of the leading 
twist pieces of the structure functions. The latter requires the 
knowledge of the structure functions at some scale in the whole 
interval $x\in[0,1[$. At present no such data are available for 
the chiral odd structure functions $h_T(x)$ and $h_L(x)$. Therefore 
and in anticipation of results from {\em RHIC} and or {\em HERMES} we 
apply leading order evolution procedures to evolve the structure 
function from the model scale, $Q_0^2=0.4 {\rm GeV}^2$ to 
$Q^2=4{\rm GeV}^2$. In Figs. \ref{fig_ht2p}a and \ref{fig_ht2p}b we 
display the results of the two step process of projection and evolution 
for the twist--2 transverse structure function, $h_T^{p}(x,Q^2)$ and 
$h_L^{p(2)}(x,Q^2)$, respectively for a constituent quark mass
of $m=400 {\rm MeV}$. 
In figure \ref{fig_h2bllp} we present the evolution of 
$h_L^{p}(x)$ along with its decomposition into terms of the leading 
twist--2 contribution, $2x \int_{x}^1\ dy h^p_T(y,Q^2)/y^2$, and the 
remaining twist--3 piece, $\overline{h}^p_L(x,Q^2)$.
As in the case of the polarized structure
function, $g_2(x,Q^2)$, the non--trivial twist--3
piece arises as a result of the binding of the constituent 
quarks through the pion fields acting as effective non--perturbative 
gluonic modes. The twist--3 contribution is evolved according to the 
large $N_C$ scheme \cite{Bal96,Ali91,Io95} outlined in the preceding
section (and in Appendix C).  Similarly in Figs. \ref{fig_ht2n} and 
\ref{fig_h2blln} we display the projection and 
evolution procedure to the twist--2 and 3 contribution to the neutron 
structure functions, $h_L^{n(2)}(x,Q^2)$ and $\overline{h}_L^n(x,Q^2)$,
respectively.
\begin{figure}[ht]
\centerline{
\epsfig{figure=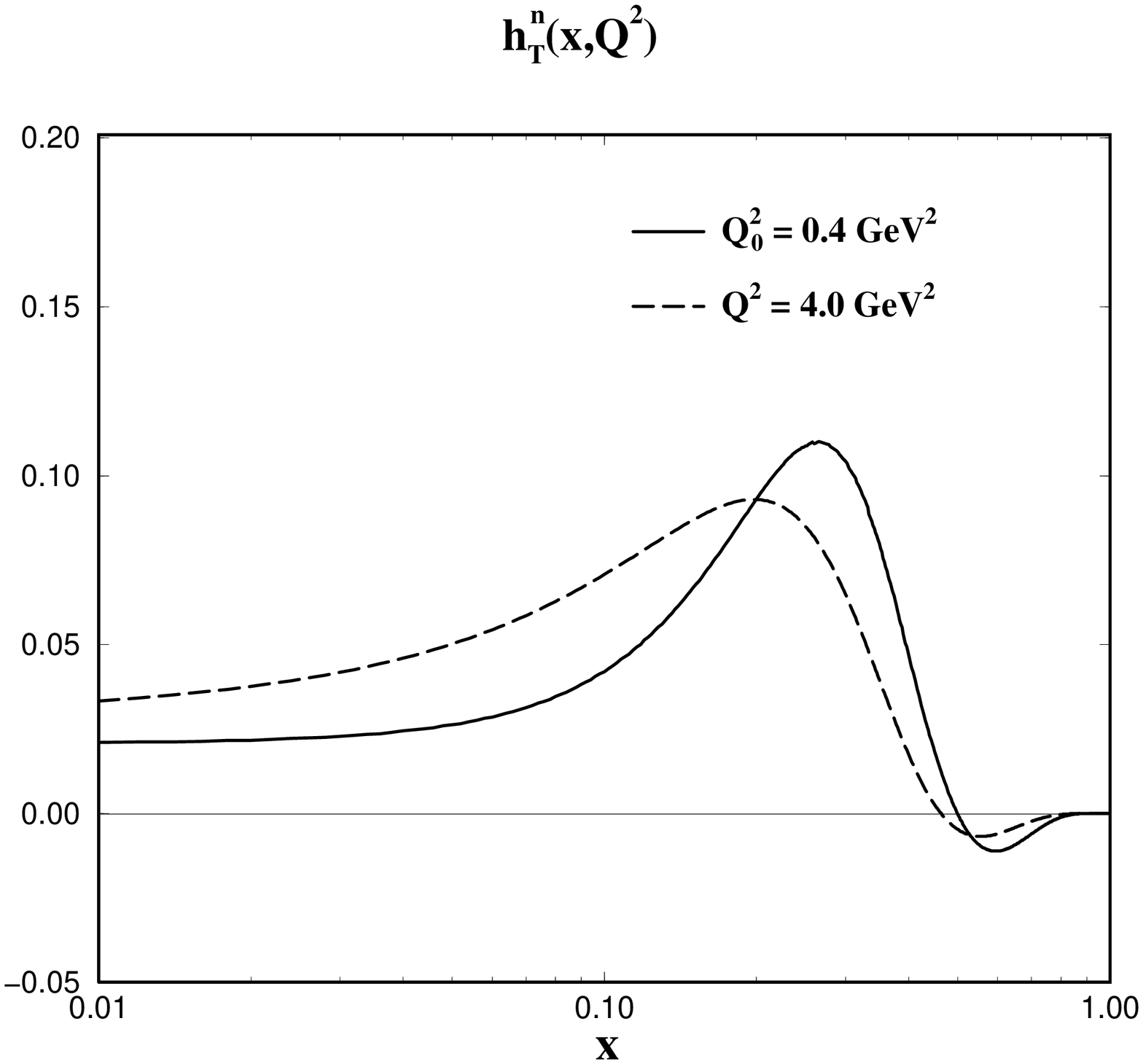,height=8.5cm,width=8.0cm,angle=270}
\hspace{-0.5cm}
\epsfig{figure=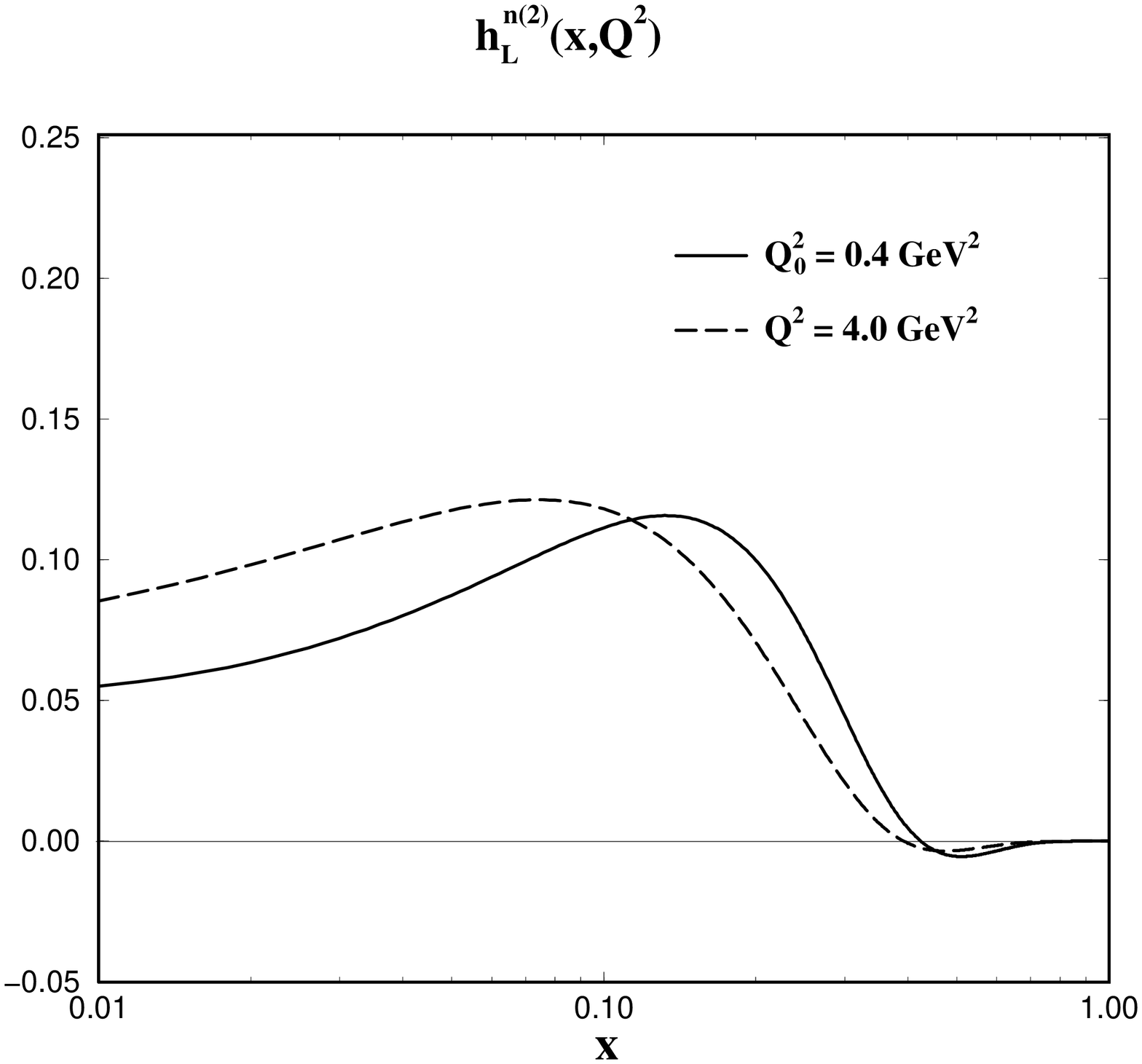,height=8.5cm,width=8.0cm,angle=270}}
\caption{\label{fig_ht2n}
Left panel:The evolution
of $h_{T}^{n}\left(x ,Q^{2}\right)$
from $Q^2_0=0.4 {\rm GeV}^2$ (solid line) to
$Q^2=4 {\rm GeV}^2$ (long--dashed line) for the constituent
quark mass $m=400 {\rm MeV}$.
Right panel: The evolution
of the twist--2 contribution to the longitudinal chiral odd
structure function,
$h_{L}^{n(2)}\left(x ,Q^{2}\right)$
from $Q^2_0=0.4 {\rm GeV}^2$ (solid line) to
$Q^2=4 {\rm GeV}^2$ (long--dashed line) for $m=400 {\rm MeV}$.}
\end{figure}
\begin{figure}[ht]
\centerline{
\epsfig{figure=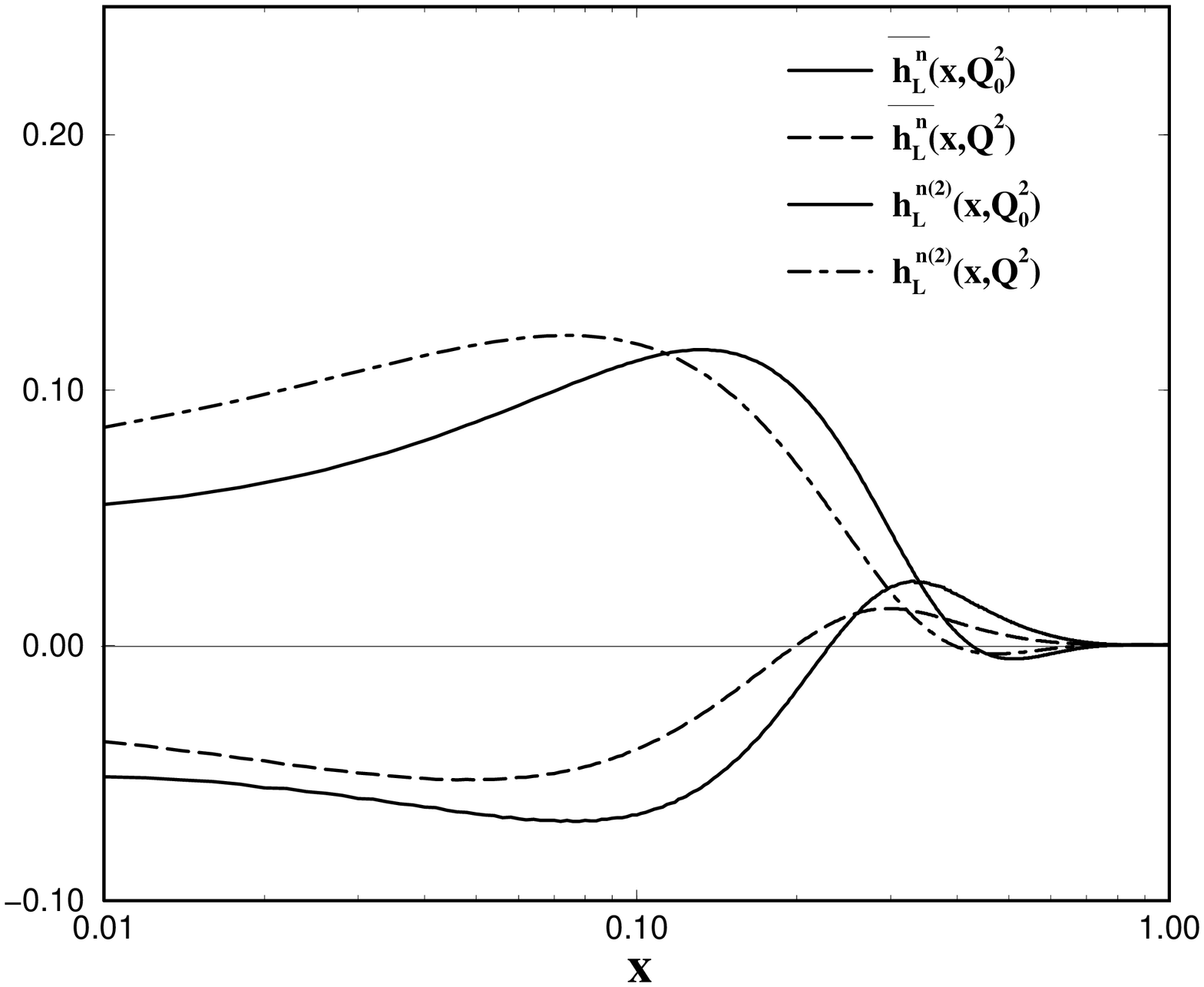,height=8.5cm,width=8.0cm,angle=270}
\hspace{-0.5cm}
\epsfig{figure=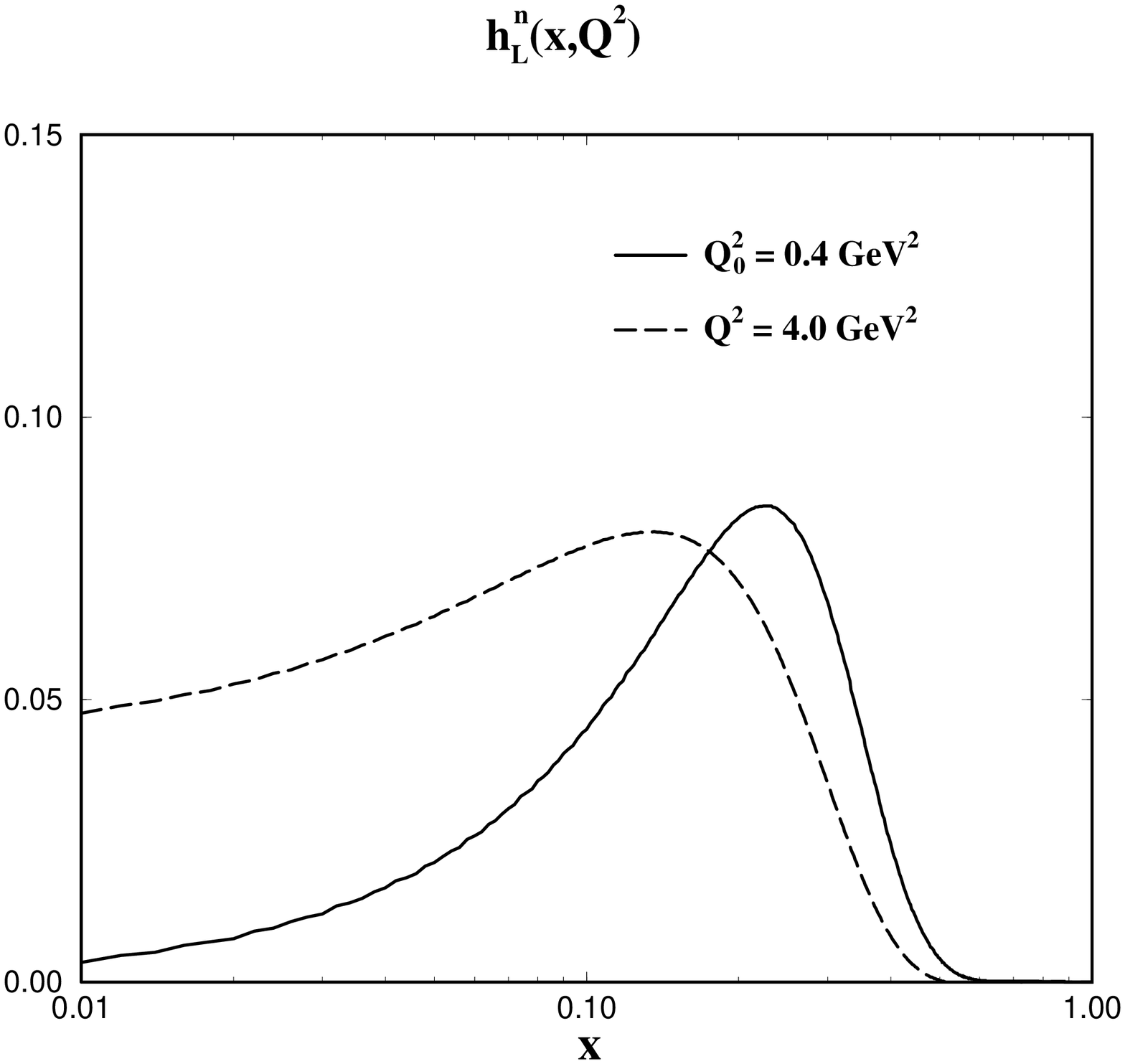,height=8.5cm,width=8.0cm,angle=270}}
\caption{\label{fig_h2blln}
Left panel:  The evolution of the twist--3
contribution to the longitudinal chiral odd
structure function, $\overline{h}_L^n(x,Q^2)$
along with the corresponding twist--2 piece,
$h_{L}^{n(2)}\left(x ,Q^{2}\right)$.
Right panel: The evolution
of $h_{L}^{n}\left(x ,Q^{2}\right)=h_{L}^{n(2)}\left(x ,Q^{2}\right)
+\overline{h}_L^n(x,Q^2)$
from $Q^2_0=0.4 {\rm GeV}^2$ (solid line) to
$Q^2=4 {\rm GeV}^2$ (long--dashed line) for the constituent
quark mass $m=400 {\rm MeV}$.}
\end{figure}

Besides the absolute magnitudes, the major difference between the chiral 
odd structure functions of the proton and the neutron is that the latter 
drop to zero at a lower value of $x$. As can be observed from figure 
\ref{fig_htnp} this is inherited from the model chiral odd structure 
function at the low momentum scale and can be linked to the smallness of 
the down quark component of $h_T$, {\it cf.} figure \ref{fig_htud}. 
Apparently the projection and evolution program does not alter this 
picture.

We would also like to compare our results from the NJL chiral soliton
model to those obtained in other approaches. A MIT bag model calculation of 
the isovector contribution $6(h_T^{p}-h_T^{n})$ has been presented 
in ref \cite{Ja92}. In shape ({\it e.g.} position of the maximum) that 
result is quite similar to ours. However, the absolute value is a bit 
larger in the MIT bag model. This reflects the fact that in the MIT bag 
model the isovector combinations of the axial and tensor charges turn 
out to be bigger than in the present model. Additionally, the QCD evolution 
of the MIT bag model prediction for $h_T$ has been studied in 
ref \cite{St93} utilizing the Peierls--Yoccoz projection as in ref \cite{Sc91}. 
In that case the maximum at $x\approx0.5$ gets shifted to a value as low 
as  $x=0.2$. Also the structure function becomes rather broad at the large 
scale.  The fact that in that calculation the evolution effects are more 
pronounced than in the present approach is caused by the significantly 
lower scale ($\mu_{\rm bag}=0.08{\rm GeV}^2$) used in  ref \cite{St93}.
On the other hand our results 
are quite different to those obtained in the QCD sum rule approach of 
ref \cite{Io95}. The sum rule approach essentially predicts $h_T$ to be 
constant in the interval $0.3<x<0.8$. For small values of $x$ the authors 
of ref \cite{Io95} assume a Regge behavior. In the (covariant) constituent 
quark model of Suzuki and Shijetamin \cite{Su97} a result similar to ours 
is obtained when effects attributed to Goldstone bosons are included. 
Otherwise the maximum of their distribution is about 50\% larger than in our
calculation. These authors also observe that in magnitude the down quark 
component is significantly smaller than the up quark piece. The chiral 
chromo--dielectric model of Barone et al. \cite{Ba97} predicts a similar 
shape for $h_T$ but their distribution $h_T^{(u)}$ is larger than the one 
in NJL chiral soliton model. This is also reflected by the sizable value 
for the isovector tensor charge $\Gamma_T^V(Q^2=25{\rm GeV}^2)=1.22$ and 
$\Gamma_T^V(Q^2=0.16{\rm GeV}^2)=1.53$ in that approach. 
In the Isgur Karl model (which has $h_1=g_1$) the maximal value of $h_1$
is only about have as big as in our calculation, {\it cf.} figure 1
in ref \cite{Sc97a}.

\begin{figure}[ht]
\centerline{
\epsfig{figure=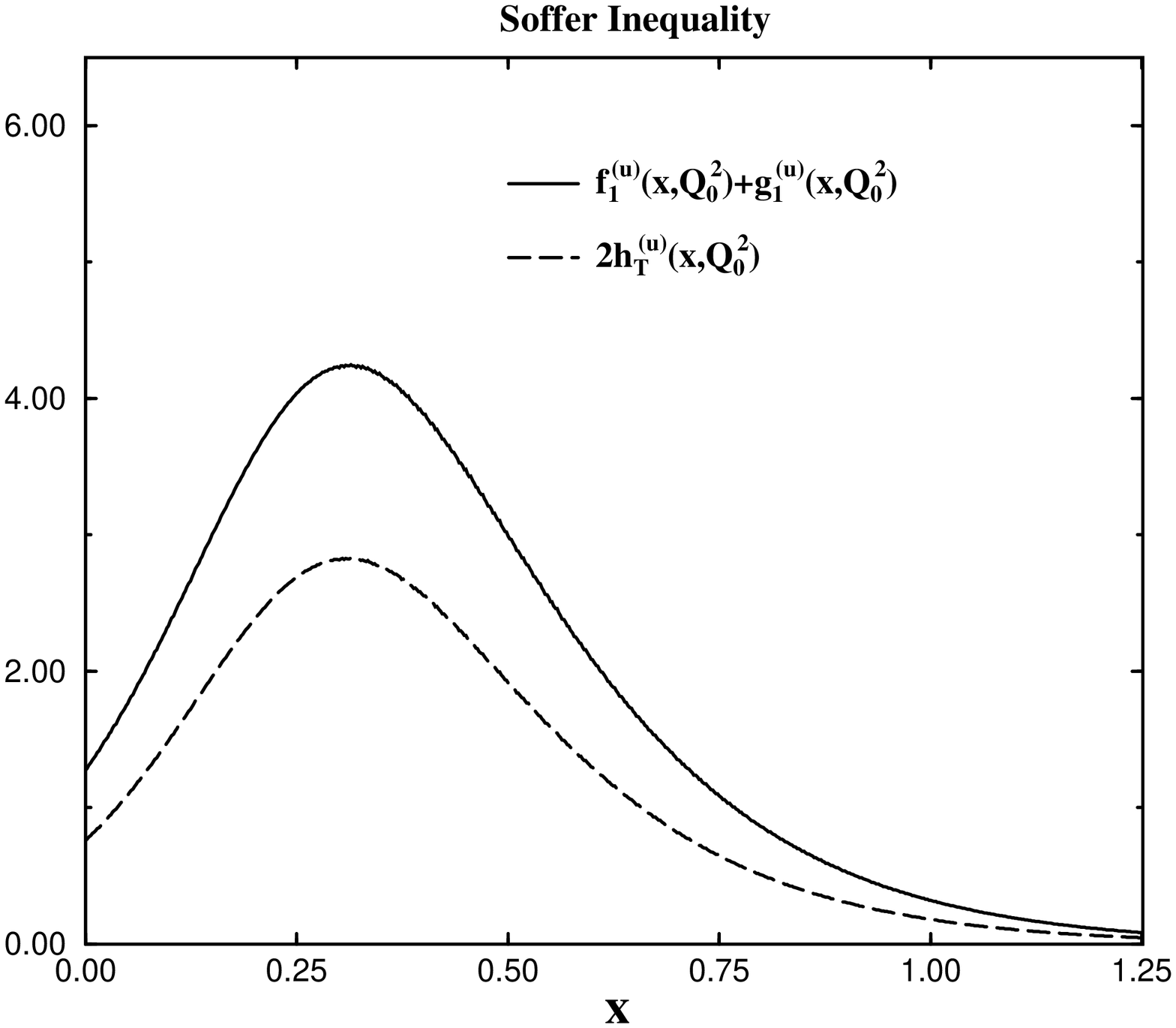,height=8.5cm,width=8.0cm,angle=270}
\hspace{-0.5cm}
\epsfig{figure=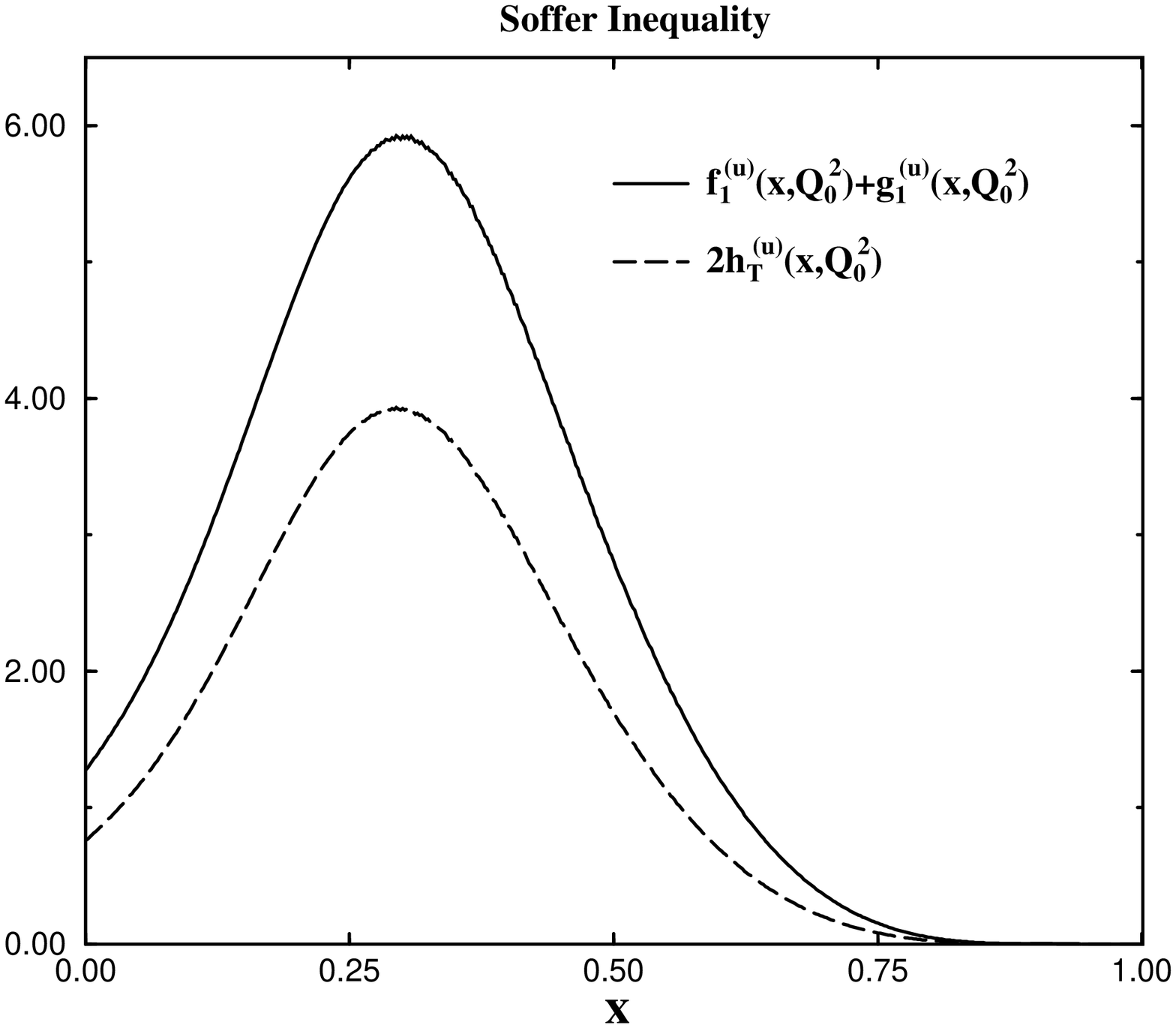,height=8.5cm,width=8.0cm,angle=270}}
\caption{\label{fig_sofu}
Left panel (\protect\ref{fig_sofu}a): 
The Soffer inequality for the chiral even 
combination $f_1^{(u)}(x,Q_0^2)+g_1^{(u)}(x,Q_0^2)$ (solid line)
of the effective up--quark distributions
and the chiral odd structure function $2\ h_T^{(u)}(x,Q_0^2)$
(long--dashed line) for a constituent quark mass of $m=400 {\rm MeV}$,
calculated in the nucleon rest frame (RF). 
Right panel: (\protect\ref{fig_sofu}b) Same as 
figure \ref{fig_sofu}a calculated in the infinite momentum
frame (IMF). The transformation prescription is given in
eq (\protect\ref{fboost}).}
\end{figure}

\begin{figure}[ht]
\centerline{
\epsfig{figure=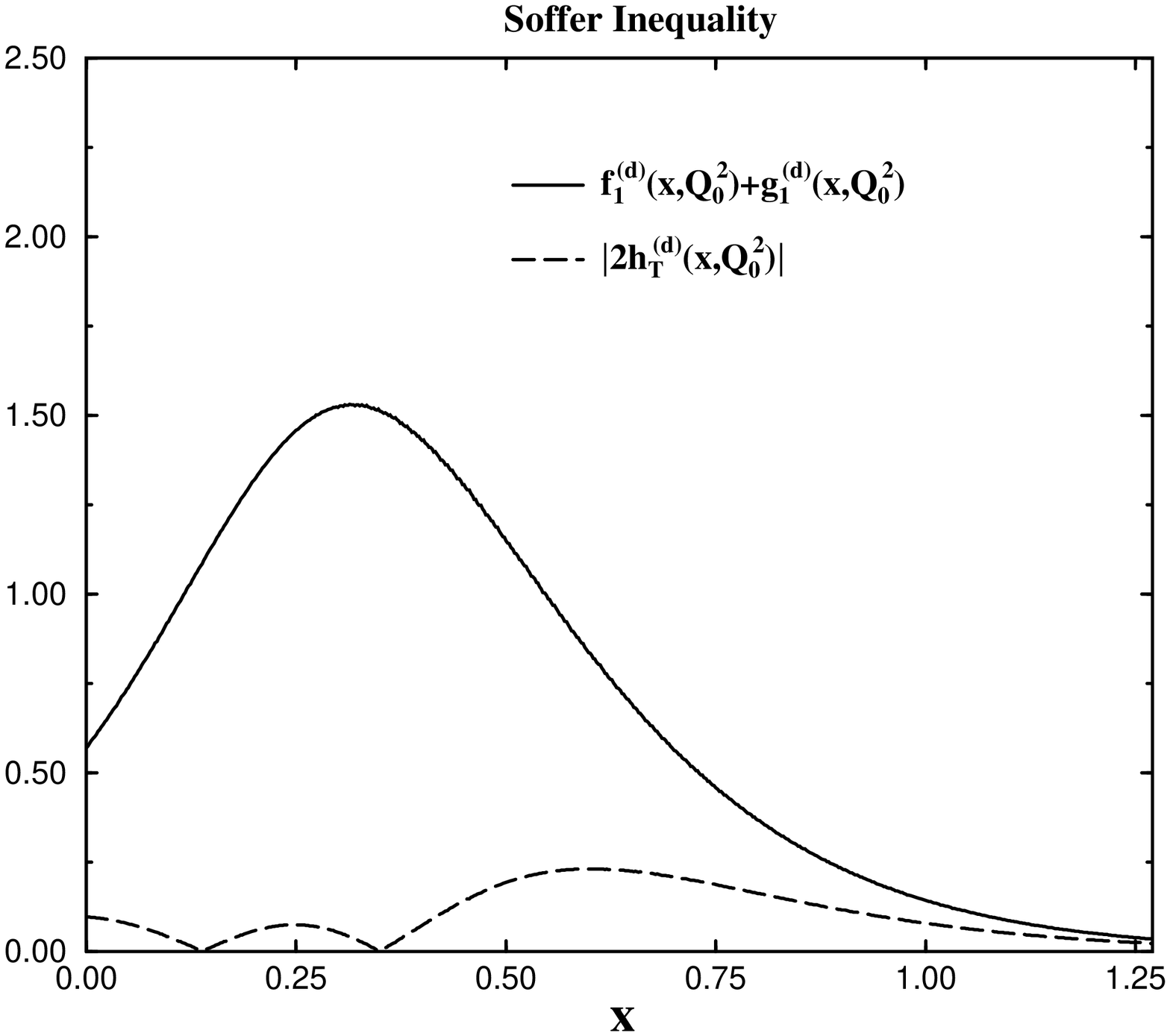,height=8.5cm,width=8.0cm,angle=270}
\hspace{-0.5cm}
\epsfig{figure=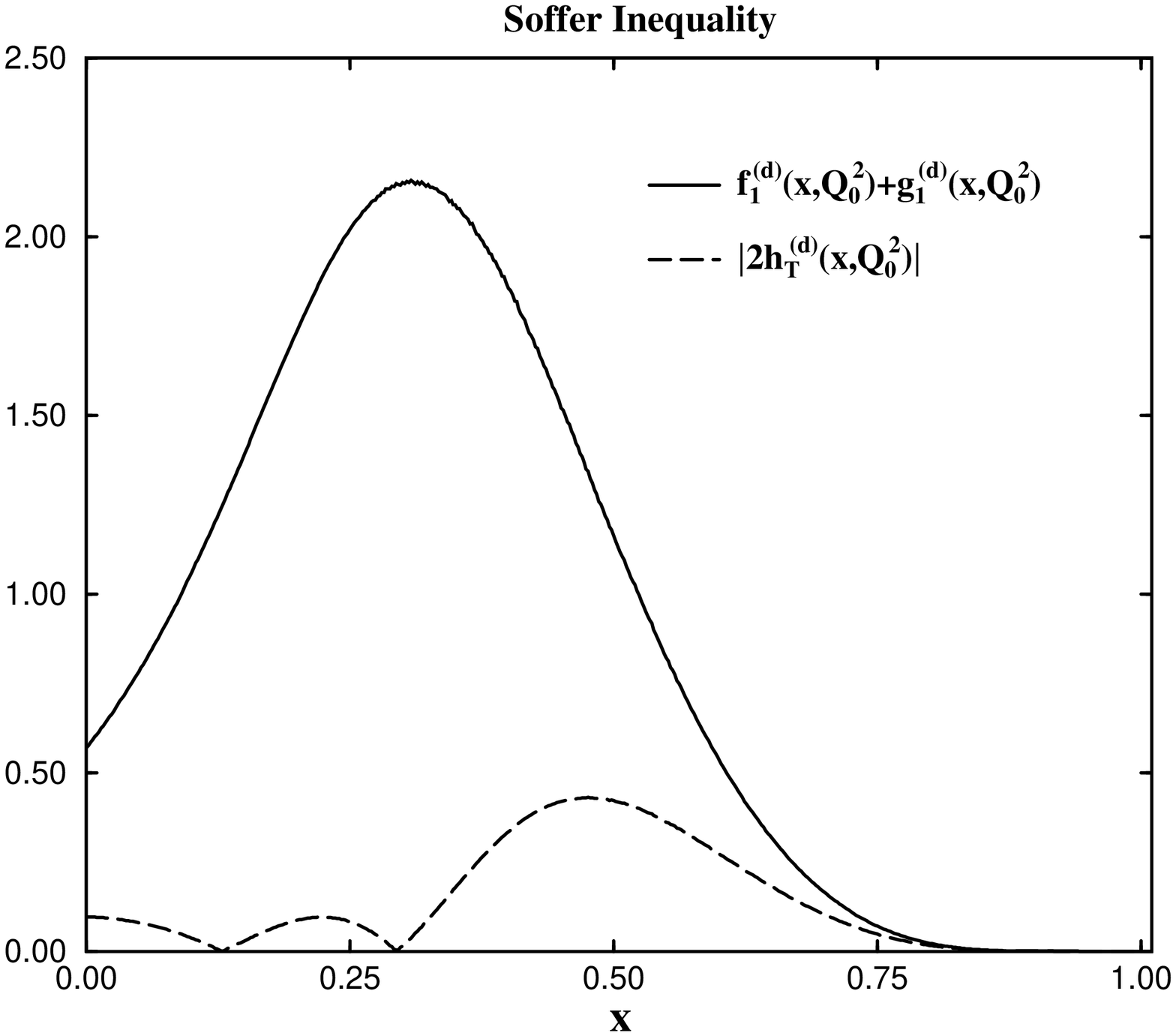,height=8.5cm,width=8.0cm,angle=270}}
\caption{\label{fig_sofd}
Same as figure \protect\ref{fig_sofu} for the down quark combination.}
\end{figure}

For completeness we also demonstrate in figs \ref{fig_sofu} and 
\ref{fig_sofd} that at the 
low model scale $Q_0^2=0.4{\rm GeV^2}$ Soffer's inequality \cite{So95} is 
satisfied. This inequality relates the nucleon chiral odd distribution
functions to both the unpolarized $f_1^{(q)}(x,Q_0^2)$ and polarized 
$g_1^{(q)}(x,Q_0^2)$ structure functions
\be
f_1^{(q)}(x,Q^2)+g_1^{(q)}(x,Q^2)\ge 2\ | h_T^{(q)}(x,Q^2) | \ .
\label{sofineq}
\ee
Here the superscript refers to the flavor combination which projects 
onto up and down quark quantum numbers ($q=u,d$). 
Note again, that this projection refers
to the constituent quarks which contain some non--perturbative 
gluonic distributions. 
In figure \ref{fig_sofd} we display the down quark component 
of the inequality (\ref{sofineq}). As this component of $h_T$ is almost
negligible the inequality is satisfied by the unpolarized structure
function $f_1^{(d)}$ being larger in magnitude than the polarized one 
$g_1^{(d)}$.  For the constituent quark mass $m=450{\rm MeV}$ we also find 
that Soffer's inequality is satisfied at the model scale $Q_0$. The 
only remarkable difference to $m=400{\rm MeV}$ is that the 
down quark component of $h_T^{(d)}$ does not possess nodes. This can already
be inferred from figures \ref{fig_htud} and \ref{fig_htudpr}.

A more thorough model study of Soffer's inequality for scales other 
than $Q_0^2$ which also contains the next--to--leading order 
contributions in the evolution program is subject to further 
investigations \cite{Ga98}. The next--to--leading order calculation 
for $h_T$ \cite{Sc97a} in the Isgur--Karl model indicates
that its scale dependence is slightly mitigated by the inclusion
of next--to--leading order contributions.

\bigskip
\section{Conclusions}
\bigskip

In this paper we have presented the NJL chiral soliton model calculation 
of the leading twist parts of the transverse and longitudinal chiral odd 
structure (distribution) functions of the nucleon. Data on these 
distribution functions should eventually be available from DIS experiments 
in the fragmentation regions (in conjunction with fragmentation functions) 
or be extracted from Drell--Yan experiments. These structure 
functions serve to complete our picture of the spin distributions of 
the nucleon. The most important feature of the present quark based model 
is that it is chirally invariant and that this symmetry is dynamically
broken. After bosonization the NJL model becomes an effective meson theory 
in which baryons emerge as self--consistent soliton solutions exactly the 
way as expected from large $N_C$ considerations in QCD. Chiral soliton 
models are particularly interesting in the context of the nucleon spin 
structure as these models nicely explain the small contribution of 
the quarks the total nucleon spin. 

In the NJL chiral soliton model there are two contributions to nucleon 
properties.  First, there is the contribution of the distinct valence 
quark level. This is the lowest level in the quark spectrum and bound in 
the background of the chiral soliton.  Second, there is the part which 
is associated to the polarization (by the soliton) of the vacuum. For 
many static nucleon properties the latter contribution is quite small,
in particular for those which are related to the axial (spin) properties 
of the nucleon. This is a strong indication that the vacuum contribution
to the chiral odd structure functions is negligible as well. Hence 
it seems more important to include substantial $1/N_C$ corrections
to the valence quark contribution. These corrections come about when 
projecting the soliton onto states with good spin and isospin, {\it i.e.} 
proton and neutron. Inclusion of these $1/N_C$ corrections together with 
a consistently regularized treatment of the vacuum polarization is 
technically rather involved and beyond the scope of the present paper.
The numerical results for the tensor charge indicate that the 
vacuum contributions are in fact negligibly small. 

When the model structure functions are computed one immediately 
recognizes that they have improper support due to the breaking 
of translational invariance by the background soliton, {\it i.e.} 
the structure functions do not exactly vanish for $x>1$. This 
can be cured by Lorentz boosting to the infinite momentum frame which 
is particularly suited for DIS processes. Although the un--boosted
structure functions are negligibly small at $x>1$ the transformation 
to this frame is essential and has sizable effects on the structure
functions at moderate $x$. However, the most important 
issue when comparing the model predictions to (not yet available)
experimental data is the observation that the model represents 
QCD at a low momentum scale $Q_0^2$. {\it A priori} this scale 
represents an additional parameter to the model calculation 
which, for consistency, has to be smaller than the ultraviolet 
cut--off of the model $\Lambda^2=0.56{\rm GeV}^2$. For the model 
under consideration we previously fixed $Q_0^2$ when studying the 
unpolarized structure functions and found $Q_0^2=0.4{\rm GeV}^2$. 
The important logarithmic corrections
to the model structure functions are then obtained within a generalized
GLAP evolution program. In this context we have restricted ourselves
to a leading order (in $\alpha_{\rm QCD}$) calculation because 
the anomalous dimensions, which govern the QCD evolution, for the 
twist--3 piece of the longitudinal part of the chiral odd structure 
are only known to that order. As the full evolution to the longitudinal
structure function involves both twist--2 and twist--3 pieces this 
restriction is consistent. We have seen that the QCD evolution of the 
chiral odd structure function leads to sizable enhancements at low $x$, 
{\it i.e.} in the region $0.01\le x\le 0.10$. In this respect the present 
situation is similar to that for the polarized structure functions.
A difference to the polarized structure function is that the lowest moment
is not protected against logarithmic corrections, even at leading order
in $\alpha_{\rm QCD}$. For the nucleon tensor charge we thus find a 
reduction of about 10\% upon evolution to $Q^2=4.0{\rm GeV}^2$.
We have also compared the neutron and proton chiral odd structure 
functions. This has been achieved by the inclusion of the $1/N_C$
cranking corrections. In absolute value the proton structure functions
are about twice as large as those of the neutron. Furthermore the 
neutron structure functions drop to zero at a lower value of $x$.
These two effects can be linked to the down quark component of the 
transverse nucleon chiral odd distribution functions being significantly 
smaller than the component with up--quark quantum numbers. We have also 
observed that neither of these features is effected by the evolution 
program.

\bigskip
\section*{Acknowledgements}
\bigskip
This work has been supported in part by the
Deutsche Forschungsgemeinschaft (DFG) under contract Re 856/2-3,
and by the U.S. Department of Energy (D.O.E.) under
contract DE--FE--02--95ER40923.
One of us (LG) is grateful to G. R. Goldstein for helpful comments
and to K. A. Milton for encouragement and support.

\appendix

\bigskip
\section{Bilocal Light Cone Distributions}
\bigskip

In this appendix we outline the steps giving rise to 
the starting point of our calculation, eqs (\ref{cht}) and 
(\ref{chl}) in section 3.  

It is well known that the chiral odd spin--dependent structure 
functions $h_T(x)$ and $h_L(x)$ do not contribute to the 
hadronic tensor in DIS.  $h_T(x)$ was first studied by Ralston 
and Soper \cite{Ra79} in the context of polarized Drell--Yan 
processes while $h_L(x)$ was more recently detailed by Jaffe and 
Ji \cite{Ja92}. In the latter study a general Lorentz decomposition 
of invariant matrix elements of the characteristic bilocal operators 
in hard processes, $\overline{\Psi}(0)\Gamma_{\mu}\Psi(\lambda n)$ 
was performed.\footnote{See ref \cite{Io95} for the definition of
chiral odd structure functions in the language of a hadronic
tensor. Actually such a definition is sufficient to carry over the 
QCD definition of the chiral odd structure functions to the NJL model 
because formally the NJL model currents are identically to those 
in QCD.} Adopting light--cone variables reveals that up to twist--3 
there are six invariant structure functions that characterize the
nucleon. The leading twist (two and three) spin--dependent 
contributions  emerging from the Lorentz decomposition of these 
hard processes are given by\cite{Ja96},
\be
\int\, \frac{d\lambda}{2\pi}\, e^{i\lambda x}\langle P,\, S|
\overline{\Psi}(0)i\sigma_{\mu\nu}\gamma_5\Psi
\left(\lambda \, n\right)| P,\, S\rangle 
&=& 2\left\{ h_T(x)\left(S_{\perp \mu}p_{\nu}
-S_{\perp \nu}p_{\mu}\right)/\, M
\right.
\nonumber \\ && \hspace{.2cm}
\left. 
+\, h_L(x)M\left(p_{\mu}p_{\nu}\, -\, p_{\nu}p_{\mu}\right)\, 
S\cdot n\right\}\, ,
\label{chodd}
\ee
\be
\int\, \frac{d\lambda}{2\pi}\, e^{i\lambda x}\langle p,\, S|
\overline{\Psi}(0)i\gamma_{\mu}\gamma_5\Psi
\left(\lambda \, n\right)| p,\, S\rangle 
&=& 2\left\{ g_L(x)p_{\mu}S\cdot n + g_T(x)S_{\perp \mu}\right\}
\label{cheven}
\ee
where $p^{\mu}$ and $n^{\mu}$ define a light-like coordinate
system, {\it i.e.} $p\cdot n=1$, $p^2=n^2=0$ and $|PS\rangle$ 
denotes the nucleon state with four momentum $P$ and spin $S$.
In the system where the nucleon is moving along the 
$\bbox{\hat{z}}$ direction one conveniently defines the four vectors 
\be
p^{\mu}&=&\frac{{\cal P}}{\sqrt{2}}(1,0,0,1)
\nonumber \\  
n^{\mu}&=&\frac{1}{\sqrt{2}}{\cal P}(1,0,0,-1)\ .
\ee
In this system the nucleon momentum is given by
$P^\mu =p^\mu + Mn^\mu/2$ and spin $S^\mu$ is decomposed 
as $S^\mu=(S\cdot n) p^\mu + (S\cdot p) n^\mu + 
S_\perp^\mu$.  Finally, ${\cal P}\to \infty$ corresponds to the 
infinite momentum frame (IMF) and ${\cal P}=M/\sqrt{2}$ corresponds 
to the nucleon rest frame (RF). Utilizing the convenient 
projection properties of the light--like vectors the defining 
equation (\ref{chodd}) may be inverted.
One obtains for the chiral odd structure functions
\be
h_T(x)&=&\frac{1}{M}\int\frac{d\lambda}{2\pi}
e^{i\lambda x}
\langle P S_{\perp}|\Psi_+^{\dagger}(0)\gamma_{\perp}\gamma_5
\Psi_+(\lambda n)| P, S_{\perp}\rangle
\ee
and
\be
h_L(x)&=&\frac{1}{2M}\int\frac{d\lambda}{2\pi}
e^{i\lambda x}
\langle P S_{z}|\Psi_-^{\dagger}(0)\gamma_0\gamma_5\Psi_+(\lambda n)
\nonumber \\ && \hspace{2.5cm}
-\Psi_+^{\dagger}(0)\gamma_0\gamma_5\Psi_-(\lambda n)| P, S_{z}\rangle\ .
\ee
The quark bilocals describe the propagation of the intermediate
constituent quark which is struck by the external source. The
forward propagation is described by $x\ge0$ while negative
$x$ parameterize an intermediate quark which moves backward.
In what follows we will only consider positive $x$ in conjunction
with the contribution associated with the forward propagating quark 
$h^{(+)}(x)$. The backward contribution can easily be obtained from 
$h^{(+)}(-x)$. Finally, noting the change of variables from light--like
coordinates $(\eta,\lambda,\bbox{\xi}_{\perp})$ 
to light--cone coordinates $(\xi^+,\xi^-,\bbox{\xi}_{\perp})$ 
where in particular,
\be
\xi^+&=&\eta{\cal P}\quad {\rm and}\quad
\xi^-=\frac{\lambda}{\cal P}
\ee
yields the chiral odd quark transverse distributions,
\be
h^{(+)}_T(x)&=&\frac{\sqrt{2}}{4\pi}\ \int d\xi^-
{\rm exp}(-i\xi^-\frac{Mx}{\sqrt{2}})
\nonumber \\ && \hspace{1.5cm}
\times\
\langle \bbox{S}_{\perp}|\Psi_+^\dagger(\xi) 
\gamma_\perp\gamma_5\Psi_+(0)
|\bbox{S}_\perp\Big\rangle_{\xi^+=\bfxi_\perp=0}\ ,
\label{chtq}
\ee
and the longitudinal contribution,
\be
h^{(+)}_L(x)&=&\frac{\sqrt{2}}{8\pi}\ \int  d\xi^-
{\rm exp}(-i\xi^-\frac{Mx}{\sqrt{2}})
\nonumber \\ && \hspace{1.0cm}
\times 
\langle \bbox{S}_z|\Psi_+^\dagger(\xi) 
\gamma_0\gamma_5\Psi_-(0)
-\Psi_-^\dagger(\xi) \gamma_0\gamma_5{\cal Q}^2\Psi_+(0)| 
\bbox{S}_z\rangle_{\xi^+=\bfxi_\perp=0}\ .
\label{chlq}
\ee
These equations represent the starting point of section 3.

\bigskip
\section{Chiral Odd Structure Functions in the 
NJL Soliton Model}
\bigskip

In this appendix we derive and summarize the explicit expressions for 
the chiral odd structure functions, eqs (\ref{hltnjl}). The first step
is to construct the eigenfunctions of the single particle Dirac 
Hamiltonian (\ref{hamil}) in coordinate space. The hedgehog {\it
ansatz} (\ref{hedgehog}) connects coordinate space with isospace and
these eigenfunctions are also eigenstates of the grand spin 
operator
\be
{\mbox{\boldmath $G$}}=
{\mbox{\boldmath $J$}}+\frac{{\mbox{\boldmath $\tau$}}}{2}
={\mbox{\boldmath $l$}}+\frac{{\mbox{\boldmath $\sigma$}}}{2}
+\frac{{\mbox{\boldmath $\tau$}}}{2}
\label{gspin}
\ee
which is the sum of the total spin ${\mbox{\boldmath $J$}}$ and the
isospin ${\mbox{\boldmath $\tau$}}/2$. The spin itself is decomposed
into orbital angular momentum ${\mbox{\boldmath $l$}}$ and intrinsic
spin ${\mbox{\boldmath $\sigma$}}/2$. Denoting by $M$ the grand 
spin projection quantum number the tensor spherical harmonics which 
are associated with the grand spin may be written as
${\cal Y}^{G,M}_{l,j}({\hat {\mbox{\boldmath $r$}}})$. Note that 
these tensor spherical harmonics are two--component spinors in 
both spin and isospin spaces. Given a  profile function 
$\Theta(r)$ the numerical diagonalization of the Dirac Hamiltonian 
(\ref{hamil}) yields the radial functions 
$g_\mu^{(G,+,1)}(r),f_\mu^{(G,+,1)}(r)$, etc. in the decomposition
({\it cf.} ref \cite{Ka84})
\be
\Psi_\mu^{(G,+)}(\mbox{\boldmath $r$})=
\pmatrix{ig_\mu^{(G,+;1)}(r){\cal Y}^{G,M}_{G,G+\frac{1}{2}} 
({\hat {\mbox{\boldmath $r$}}}) \cr
f_\mu^{(G,+;1)}(r) {\cal Y}^{G,M}_{G+1,G+\frac{1}{2}} 
({\hat {\mbox{\boldmath $r$}}})\cr} +
\pmatrix{ig_\mu^{(G,+;2)}(r){\cal Y}^{G,M}_{G,G-\frac{1}{2}} 
({\hat {\mbox{\boldmath $r$}}})\cr
-f_\mu^{(G,+;2)}(r) {\cal Y}^{G,M}_{G-1,G-\frac{1}{2}}
({\hat {\mbox{\boldmath $r$}}})\cr}
\label{psipos} \\ \nonumber \\
\Psi_\mu^{(G,-)}(\mbox{\boldmath $r$})=
\pmatrix{ig_\mu^{(G,-;1)}(r) {\cal Y}^{G,M}_{G+1,G+\frac{1}{2}} 
({\hat {\mbox{\boldmath $r$}}})\cr
-f_\mu^{(G,-;1)}(r) {\cal Y}^{G,M}_{G,G+\frac{1}{2}} 
({\hat {\mbox{\boldmath $r$}}})\cr} +
\pmatrix{ig_\mu^{(G,-;2)}(r){\cal Y}^{G,M}_{G-1,G-\frac{1}{2}} 
({\hat {\mbox{\boldmath $r$}}})\cr
f_\mu^{(G,-;2)}(r) {\cal Y}^{G,M}_{G,G-\frac{1}{2}} 
({\hat {\mbox{\boldmath $r$}}})\cr}.
\label{psineg}
\ee
The second superscript ($\pm$) denotes the intrinsic parity, which 
also is a conserved quantum number.\footnote{The total parity is given 
by the product of the intrinsic parity and $(-)^G$.} Note that for 
the $G=0$ channel, which contains the mean--field contribution to the 
valence quark wave--function in eq (\ref{valrot})
\be
\Psi_{\rm v}({\mbox{\boldmath $r$}})=\left(
\ba{c}
i g_{\rm v}(r) {\cal Y}_{0,\frac{1}{2}}^{0,0} 
\left(\hat{\mbox{\boldmath $r$}}\right) \\
f_{\rm v}(r) {\cal Y}_{1,\frac{1}{2}}^{0,0}
\left(\hat{\mbox{\boldmath $r$}}\right)
\ea
\right),
\label{valcl}
\ee
only the components with $j=+1/2$ are allowed. In addition to 
this mean--field  piece (\ref{valcl}) the complete valence quark 
wave--function (\ref{valrot}) also contains the cranking correction, 
which dwells in the channel with $G=1$ and negative intrinsic parity.

The discretization ($\mu$) is accomplished by choosing suitable 
boundary conditions at a radial distance which is large compared to 
the soliton extension \cite{Ka84,We92}. This calculation yields the 
energy eigenvalues $\epsilon_\mu$, which enter the energy functional 
(\ref{efunct}). The soliton configuration is finally determined by 
self--consistently minimizing this energy functional. In ref 
\cite{Al94a} the numerical procedure is described in detail.

We continue by making explicit the Fourier transform of 
eq (\ref{valrot}),
\be
\tilde{\psi}_{\rm v}\left(\mbox{\boldmath $p$}\right)=
\int \frac{d^3x}{4\pi}\psi_{\rm v}\left(\mbox{\boldmath $x$}\right)
{\rm exp}\left(i\mbox{\boldmath $p$}\cdot
\mbox{\boldmath $x$}\right)=
\tilde{\Psi}_{\rm v}\left(\mbox{\boldmath $p$}\right)+Q_\mu
\tilde{\Psi}_{\mu}\left(\mbox{\boldmath $p$}\right)\ .
\label{ft0}
\ee
The leading order in $N_C$ valence quark contribution is just the
Fourier transform of (\ref{valcl})
\be
\tilde\Psi_{\rm v}({\mbox{\boldmath $p$}})=i\left(\ba{c}
\gvp\ {\cal Y}_{0,\frac{1}{2}}^{0,0} \left(\hat{\mbox{\boldmath $p$}}\right)
\\ \fvp\ {\cal Y}_{1,\frac{1}{2}}^{0,0}\left(\hat{\mbox{\boldmath $p$}}\right)
\ea \right)
\label{ft1}
\ee
and the cranking correction involves the Fourier transform of 
spinor with $G=1$ and negative intrinsic parity
\be
\tilde\Psi_{\mu}({\mbox{\boldmath $p$}})=
-i\left(\ba{c}\guip\ {\cal Y}_{2,\frac{3}{2}}^{1,M}
\left(\hat{\mbox{\boldmath $p$}}\right) -
\guiip\ {\cal Y}_{0,\frac{1}{2}}^{1,M}
\left(\hat{\mbox{\boldmath $p$}}\right)\\ 
\fuip\ {\cal Y}_{1,\frac{3}{2}}^{1,M}
\left(\hat{\mbox{\boldmath $p$}}\right)-
\fuiip\ {\cal Y}_{1,\frac{1}{2}}^{1,M}
\left(\hat{\mbox{\boldmath $p$}}\right)
\ea \right) \ .
\label{ft2}
\ee
Here ${\cal Y}_{l,j}^{G,M}\left(\hat{\mbox{\boldmath $p$}}\right)$ 
are the Fourier transforms of the tensor spherical harmonics 
associated with the grand spin operator (\ref{gspin}). 
The Fourier transform for the radial functions in eqs 
(\ref{ft1}) and (\ref{ft2}) is defined by
\be
\tilde{\phi}_\mu(p)=\int_{0}^{R}dr\ r^2
j_{l}(pr)\phi_{\mu}(r)\ .
\ee
Here the index $l$ of the spherical Bessel function denotes 
the orbital angular momentum of the associated tensor spherical 
harmonic. We have suppressed the grand spin index on the transforms 
of the radial wave functions for convenience. For purposes of 
notation we have also introduced the quantity $Q_\mu$ in eq 
(\ref{ft0}) which parameterizes the cranking corrections in 
eq (\ref{valrot})
\be
\frac{\langle \mu |\mbox{\boldmath $\tau$}\cdot
\mbox{\boldmath $\Omega$}|{\rm v}\rangle}
{\epsilon_{\rm v}-\epsilon_\mu}&=&\alpha^2
Q_\mu\ \left\{\frac{\delta_{M,1}}{\sqrt 2}
\left(\Omega_{1}+i\Omega_{2}\right)
-\frac{\delta_{M,-1}}{\sqrt 2}\left(\Omega_{1}-i\Omega_{2}\right)
-\delta_{M,0}\Omega_{0}\right\}\delta_{G_\mu ,1}\\ 
&=&Q_\mu\ \left\{\frac{\delta_{M,1}}{\sqrt 2}
\left(J_{1}+iJ_{2}\right)
-\frac{\delta_{M,-1}}{\sqrt 2}\left(J_{1}-iJ_{2}\right)
-\delta_{M,0}J_{0}\right\}\delta_{G_\mu ,1}
\label{defqm}
\ee
where
\be
Q_\mu \equiv\frac{1}{\alpha^2
\left(\epsilon_{\rm v}-\epsilon_\mu\right)}
\int \ dr\ r^2 \left\{\gvr\guiir\ +\fvr\fuiir\right\}\ .
\ee
In this definition we have included the total moment of inertia 
$\alpha^2$. In the proper--time regularization of the 
NJL chiral soliton model $\alpha^2$ is given by \cite{Re89}
\be
\alpha_{\rm v}^2&=&\frac{N_C}{2} \sum_{\mu\ne{\rm v}}
\frac{\left|\langle {\rm v}|\tau_3|\mu\rangle\right|^2}
{\left(\epsilon_\mu-\epsilon_{\rm v}\right)} \ ,
\label{al2v} \\
\alpha_{\rm s}^2&=&\frac{N_C}{4\alpha^2}\sum_{\mu\nu}f_{\mu\nu}(\Lambda)
\langle \mu |\tau^3|\nu\rangle\langle\nu|\tau^3|\mu\rangle \ ,
\label{al2s} \\
\alpha^2&=&\frac{1}{2}\left(1+{\rm sgn}(\epsilon_{\rm val})\right)
\alpha_{\rm v}^2+\alpha_{\rm s}^2 .
\label{altot} 
\ee
The regulator function in the vacuum contribution reads
\be
f_{\mu\nu}(\Lambda)=\frac{\Lambda}{\sqrt{\pi}}
\frac{e^{-(\epsilon_\mu/\Lambda)^2}-e^{-(\epsilon_\nu/\Lambda)^2}}
{\epsilon_\nu^2-\epsilon_\mu^2}
-\frac{{\rm sgn}(\epsilon_\nu)
{\rm erfc}\left(\left|\frac{\epsilon_\nu}{\Lambda}\right|\right)
-{\rm sgn}(\epsilon_\mu)
{\rm erfc}\left(\left|\frac{\epsilon_\mu}{\Lambda}\right|\right)}
{2(\epsilon_\mu-\epsilon_\nu)} \ .
\label{freg}
\ee
The moment of inertia enters via the quantization description for 
the collective coordinates $\bbox{\Omega}\to \alpha^2 \bbox{J}$ with 
$\bbox{J}$ being the nucleon spin operator. In this quantization 
prescription we had previously restricted the moment of inertia to 
its valence quark contribution, $\alpha_{\rm v}^2$, to ensure that the 
Adler sum rule for the unpolarized structure functions is maintained in 
the valence quark approximation \cite{We96a}. For small or moderate 
constituent quark masses the valence contribution to the moment of 
inertia is about 80\% or more \cite{We92}. This is one of the reasons 
to believe that the valence quark approximation to structure functions 
is sensible. In the case of the chiral odd structure functions (as for 
the polarized ones) the valence quark approximation appears to be even
better. As we see from table \ref{tab_1} the lowest moments of these 
structure functions are saturated to about 95\% by the valence quark 
contribution. Hence it is reasonable to assume that the vacuum 
contribution to these structure functions in negligibly small. As a 
consequence the valence quark approximation with the total moment of 
inertia substituted into the quantization rule will provide a very 
reliable estimate of the chiral odd structure functions.

Together with 
$\langle N | D_{ij} | N\rangle = - (4/3) I_i J_j$ \cite{Ad83} 
the nucleon matrix elements may now easily be computed.  Here 
$\bbox{I}$ denotes the nucleon isospin. Whenever products  
of collective coordinates and operators appear which do not
commute after canonical quantization we adopt the symmetric 
ordering.  This is consistent with fundamental requirements
such as PCAC. Defining finally the following combinations 
\be
\fupq=Q_\mu \fup \qquad {\rm and} \qquad
\gupq=Q_\mu \gup\ ,
\ee
for $i=1,2$. The isoscalar(vector) contributions to the chiral odd
structure functions (\ref{hltnjl})
read
\be
h^{I=0}_{T,\pm}(x,\mu^{2})&=&N_C\frac{5M_N}{36\pi}
\int^\infty_{M_N|x_\mp|}p dp 
\nonumber \\* && \hspace{.1cm}
\times
\left\{\gvp\gip\frac{3\csii-1}{4\sqrt{2}}-\frac{1}{2}\gvp\giip
\right.
\nonumber \\* && \hspace{.1cm}
\pm\left(\gvp\fip+\fvp\gip\right)\frac{\csi}{\sqrt{8}}
\nonumber \\* && \hspace{.1cm}
\mp\left(\fvp\giip+\gvp\fiip\right)\frac{\csi}{2}
\nonumber \\* && \hspace{.1cm}
\left.
-\fvp\fip\frac{\csii -3}{4\sqrt{2}}
-\fvp\fiip\frac{\csi2}{2}
\right\}
\label{hT0}
\ee
\be
\hspace{-1cm}
h^{I=1}_{T,\pm}(x,\mu^{2})&=&N_C\frac{M_N}{36\pi}
\int^\infty_{M_N|x_\mp|}p dp 
\nonumber \\* && \hspace{.1cm}
\times\left\{\gvp^2\pm 2\gvp\fvp\csi
+\fvp^2\left(\csii\right)\right\}\ ,
\label{hT1}
\ee
\be
\hspace{-1cm}
h^{I=0}_{L,\pm}(x,\mu^{2})&=&N_C\frac{5M_N}{36\pi}
\int^\infty_{M_N|x_\mp|}p dp 
\nonumber \\* && \hspace{.1cm}
\times\left\{\pm\gvp\gip\frac{3\csii-1}{4\sqrt{2}}
\pm\frac{1}{2}\gvp\giip
\right.
\nonumber \\* && \hspace{.1cm} 
\left.
\mp\fvp\fip\frac{1+\csii}{2\sqrt{2}}
\mp\fvp\fiip\frac{2\csii-1}{2}\right\}\ ,
\label{hL0}
\ee
\be
\hspace{-1cm}
h^{I=1}_{L,\pm}(x,\mu^{2})&=&-N_C\frac{M_N}{36\pi}
\int^\infty_{M_N|x_\mp|}p dp 
\left\{\mp\gvp^2 \pm \fvp^2 \left(2\csii-1\right)\right\}\ .
\label{hL1}
\ee
which we evaluate numerically. 
Note that the angle $\theta_p^\pm$ is related to the integration
variable $p$ via
\be
{\rm cos}\theta_p^\pm = \frac{1}{p}|M_N x \pm \epsilon_{\rm v}|\ .
\ee
In ref \cite{Po96} the contribution to 
structure function $h_T$ from {\em effective} quark 
distributions\footnote{In this work it is important to note
that the  quark distributions refer to {\em constituent quarks}, 
$m_q\approx 400 {\rm MeV}$, it is thus misleading to compare 
them with the data of {\em parton} distributions from either 
Drell--Yan or DIS processes.} was calculated omitting the cranking 
corrections and adopting an external (non--self--consistent)
meson profile.
\bigskip

\bigskip
\section{Evolution of 
$\overline{\lowercase{h}}_L(x,Q^2)$}
\bigskip
 
In this appendix we outline our technique to evolve the low scale 
model prediction for the twist--3 piece $\overline{h}_L(x,Q_0^2)$ 
to the larger scale $Q^2$. This utilizes the method described in 
refs \cite{Ka97,Sch90} based on the results of ref \cite{Go79}. 
The $Q^2$ evolution of the moments 
\be
{\cal M}_n\left[ \overline{h}_L(Q^2)\right]=
\int_0^1 dx x^n \overline{h}_L(x,Q^2)
\label{defmom}
\ee
is given by
\be
{\cal M}_n\left[ \overline{h}_L(Q^2)\right]
=\left(\frac{\alpha(Q^2)}{\alpha(Q_0^2)}
\right)^{\gamma_n^h/b_0}{\cal M}_n\left[\overline{h}_L(Q_0^2)\right]
\label{mom}\ .
\ee
Here $b_0=\left(11N_C -2n_f\right)/3$ is coefficient of the leading term
in the QCD--beta function. Also, $N_C$ and $n_f$ are the number of 
colors and flavors respectively. Within the $N_C\to\infty$ approximation 
the anomalous dimensions are \cite{Bal96}
\be
\gamma_n^h= 2N_C \left(S_n+ \gamma_{\rm E} -\frac{1}{4} + 
\frac{3}{2(n+1)} \right),
\label{adm}
\ee
with $S_n=\sum_{j=1}^n(1/j)-\gamma_{\rm E}$ where
$\gamma_E = 0.577\ldots$ is the Euler constant which has been 
introduced for later convenience.

In order to find the QCD--evolution of the structure functions
one needs to invert the Mellin--transform (\ref{defmom}). This 
can be achieved by noting that the Bernstein polynomial 
\be
b^{(N,k)}(x)=(N+1) {n\choose k} x^k(1-x)^{N-k} = \frac{(N+1)!}{k!}
\sum_{l=0}^{N-k}\frac{(-1)^l x^{k+l}}{l!(N-k-l)!}\ ,
\label{bern}
\ee
has the property,
\be
\lim_{{\scriptstyle N,k\to \infty} \atop 
{\scriptstyle k/N\to x} } b^{(N,k)}(y) = \delta (x-y)
\label{delta}
\ee
for $0<x, y <1$. This enables one to express the structure 
function via its moments
\be
\overline{h}_L(x,Q^2) &=& \lim_{{\scriptstyle N,k\to \infty} \atop
{\scriptstyle k/N\to x}} \frac{(N+1)!}{k!} \sum_{l=0}^{N-k}
\frac{(-1)^l}{l!(N-k-l)!} \int^1_{0}dy\,y^{k+l}\overline{h}_L(y,Q^2)
\label{bern1}
\ee
which depend on $Q^2$ as indicated in eq (\ref{mom})
\be
\overline{h}_L(x,Q^2) &=&
\lim_{{\scriptstyle N,k\to \infty} \atop
{\scriptstyle k/N\to x} } \frac{(N+1)!}{k!} \sum_{l=0}^{N-k}
\frac{(-1)^l}{l!(N-k-l)!}L^{\gamma_{k+l}^h/b_0}
\int^1_{0}dy\,y^{k+l}\overline{h}_L(y,Q_0^2)\ .
\label{bern2}
\ee
Here $L=\alpha(Q^2)/\alpha(Q_0^2)$ denotes the ratio of the 
running coupling constants in QCD. Unfortunately, the rapid 
oscillations in the summation over 
$l$ in (\ref{bern2}) due to the factor $(-1)^l$ preclude
numerical summation of (\ref{bern2}).
Yet,  observing that the expression
$L^{\gamma_n^h/b_0}$ may be expanded as
\be
L^{\gamma_n^h/b_0}=
a(L)\sum_{i=0}\frac{{\cal C}_i(L)}{(n+p)^{i-r(L)}},
\label{expn}
\ee
where $a(L)$, and $r(L)$ are constants determined from the 
asymptotic form ($n\to\infty$) of eq (\ref{expn})
\be
r(L)=2N_C {\rm ln}(L)/b_0\quad {\rm and} \quad
a(L)={\rm exp}\left[r(L)\left(\gamma_{\rm E}-\frac{1}{4}\right)\right]\ ,
\label{asymp}
\ee
one can perform the sum to any desired accuracy.
It should be noted that $p$ remains undetermined. It may be 
varied to control the convergence of the series (\ref{expn}).
To determine the expansion coefficients ${\cal C}_i(L)$ 
we rearrange eq (\ref{expn}) to a Fourier expansion,
\be
(1-z\, p)^r\ {\rm exp}\left[\frac{1}{2n}+\frac{3}{2n+2}
-\sum_{k=1}^{\infty} 
\left(\frac{{\cal B}_{2k}}{2k\, n^{2k}}\right)\right]
=\sum_{i=0}^{\infty}\, C_{i}(r)\, z^i
\label{herb}\ .
\ee
Here $z=1/(p+n)\Leftrightarrow n=1/z-p$ and 
${\cal C}_i(L)=C_{i}(r)$. Furthermore we have utilized the 
asymptotic expansion of 
\be
S_n={\rm ln}(n)+\frac{1}{2\, n}-
\sum_{k=1}^{\infty} \left(\frac{{\cal B}_{2k}}{2k\, n^{2k}}\right)\ ,
\ee
where the  ${\cal B}_{2k}$'s are the Bernoulli numbers. Performing a 
Taylor series to eighth order in $z$ yields the following values for 
the expansion coefficients ${\cal C}_{i}(L)=C_{i}(r(L))$ (for $p=2$)
\be
\hspace{-4cm}
C_{0}(r)&=&1,  \qquad C_{1}(r)=0, \qquad \qquad
C_{2}(r)=\frac{5}{12}\, r ,
\nonumber \\ \hspace{-4cm}
C_{3}(r)&=&\frac{1}{2}\, r, \qquad
C_{4}(r)=\frac{61}{120}\, r + \frac{25}{288}\, r^2, \qquad
C_{5}(r)=\frac{1}{2}\, r+\frac{5}{24}\, r^2 ,
\nonumber \\ \hspace{-4cm}
C_{6}(r)&=&\frac{125}{252}\, r + \frac{97}{288}\, r^2
+ \frac{125}{10368}\, r^3 ,
\nonumber \\ \hspace{-4cm}
C_{7}(r)&=&\frac{1}{2}\, r + \frac{37}{80}\, r^2
+ \frac{25}{576}\, r^3 ,
\nonumber \\ \hspace{-4cm}
C_{8}(r)&=&\frac{121}{240}\, r + \frac{354341}{604800}\, r^2
\frac{665}{6912}\, r^3 + \frac{625}{497664}\, r^4\ \ ,
\label{coeffs}
\ee
which gives more than adequate convergence of the series.

Finally we may write
\be
\overline{h}_L(x,Q^2)&=&\int_x^1 \frac{dy}{y} 
b(x,y;Q^2,Q_0^2)  \overline{h}_L(y,Q_0^2),
\label{prop}
\ee
where,
\be
b(x,y;Q^2,Q_0^2) &=& a(L) \left( \frac{x}{y}\right)^{p-1}
\sum_{i=0}\left({\rm ln}\frac{y}{x}\right)^{i+\rho-1}
\frac{{\cal C}_i(L)}{\Gamma(i+\rho)}\ 
\label{kernel}
\ee
is the evolution kernel used in eq (\ref{evkern}). It has been gained 
by using the additional relation
\be
\lim_{{\scriptstyle N,k\to \infty} \atop
{\scriptstyle k/N\to x} }\frac{(N+1)!}{k!} \sum_{l=0}^{N-k}
\frac{(-1)^l}{l!(N-k-l)!}\frac{y^{k+l}}{(k+l+p)^{i+ \rho}}
=\frac{\theta(y-x)}{\Gamma(i+\rho) y}\left(\frac{x}{y}\right)^{p-1}
\left( {\rm ln}\frac{y}{x} \right)^{i+\rho-1}\ .
\nonumber
\ee
For the numerical results presented in sections 4 and 5 we have 
verified the stability of this evolution procedure by varying the 
undetermined parameter $p$ in eq (\ref{expn}). 

\bigskip
\section{Tensor charges in the NJL chiral soliton model}
\bigskip

The conventional definition of the nucleon tensor charges reads
\be
\langle N | {\bar \Psi} \sigma_{\mu\nu} \Psi | N \rangle =
\Gamma_T^S\  {\bar u}\sigma_{\mu\nu} u
\qquad {\rm and} \qquad
\langle N | {\bar \Psi} \sigma_{\mu\nu} \tau^3 \Psi | N \rangle =
\Gamma_T^V\  {\bar u}\sigma_{\mu\nu} \tau^3 u \ .
\label{defgten}
\ee
Here $N$ again denotes the nucleon state. Note that both, 
the quark wave--function $\Psi$ as well as the nucleon 
spinor $u$, are vectors in flavor space. Momentum labels have
been omitted as the charges are defined at zero momentum transfer.
Within the NJL chiral soliton model these charges can be extracted
using standard techniques \cite{Al96}: first, sources conjugated 
to the quark bilinears ${\bar \Psi} \sigma_{\mu\nu} \Psi$ and 
${\bar \Psi} \sigma_{\mu\nu} \tau^3 \Psi$ are added to the Lagrangian
(\ref{NJL}). Subsequently the bosonized and regularized action is 
expanded to linear order in both the sources and the angular velocities 
$\mbox{\boldmath $\Omega$}$ (\ref{angvel}). The coefficients
of the source terms then provide the charge operators in the 
space of the collective coordinates $A$, which are defined in 
eq (\ref{collrot}). The corresponding matrix elements can be
straightforwardly evaluated with the means provided in 
appendix B. Finally one obtains within the proper--time 
regularization
\be
\Gamma_T^S&=&\frac{N_C}{4\alpha^2}
\left(1+{\rm sgn}(\epsilon_{\rm val})\right)
\sum_{\nu\ne{\rm val}}
\frac{\langle {\rm val}|\tau^3|\nu\rangle
\langle\nu|\beta\Sigma_3\tau^3|{\rm val}\rangle}
{\epsilon_{\rm val}-\epsilon_\nu} 
\nonumber \\ && \hspace{1cm}
+\frac{N_C}{4\alpha^2}\sum_{\mu\nu}f_{\mu\nu}(\Lambda)
\langle \mu |\tau^3|\nu\rangle
\langle\nu\beta\Sigma_3\tau^3|\mu\rangle 
\label{gtens1}\\
\Gamma_T^V&=&-\frac{N_C}{6}
\left(1+{\rm sgn}(\epsilon_{\rm val})\right)
\langle {\rm val}|\beta\Sigma_3\tau^3|{\rm val}\rangle
\nonumber \\ && \hspace{1cm}
+\frac{N_C}{6}\sum_\mu \langle\mu|\beta\Sigma_3\tau^3|\mu\rangle
{\rm sgn}(\epsilon_\mu) \
{\rm erfc}\left(\left|\frac{\epsilon_\mu}{\Lambda}\right|\right) \ .
\label{gtenv1}
\ee
Here $|\mu\rangle$ denote the eigenstates of the static 
Dirac--Hamiltonian (\ref{hamil}) and $\epsilon_\mu$ are the 
corresponding eigenvalues. Again $|\rm val\rangle$ refers to the 
distinct valence quark level. The regulator function in the isoscalar 
piece (\ref{gtens1}) is identical to the one entering 
the moment of inertia, {\it cf.} eq (\ref{freg}). Those pieces 
containing the factor $\left(1+{\rm sgn}(\epsilon_{\rm val})\right)$
are the valence contributions shown separately in table \ref{tab_1}.

As noted in section 5 we have omitted $1/N_C$ suppressed 
contributions to the isovector part $\Gamma_T^V$ which in the related
case of the axial current violate PCAC.

\end{document}